\documentclass{aa}
\usepackage{graphicx}
\usepackage{xcolor}
\usepackage{subfig}   
\usepackage{txfonts}
\usepackage{xspace}
\usepackage{hyperref}

\newcommand{\fofrfour}{fofr4\xspace}

\newcommand{\fofrsix}{fofr6\xspace}
\newcommand{\symmA}{symmetron A\xspace}
\newcommand{\symmB}{symmetron B\xspace}
\newcommand{\symmC}{symmetron C\xspace}
\newcommand{\symmD}{symmetron D\xspace}

\newcommand{\Mpch}{\,{\rm Mpc}\,\ifmmode h^{-1}\else $h^{-1}$\fi}

\begin{document}

   \title{Probing modified gravity in cosmic filaments}


   \author{Alex Ho\inst{1}
          \and
          Max Gronke\inst{2}
          \and 
          Bridget Falck\inst{1}
          \and 
          David F. Mota\inst{1}
          }

   \institute{Institute Of Theoretical Astrophysics (ITA), University of Oslo,
              0315 Oslo, Norway\\
              \email{alex.ho@astro.uio.no}
         \and
             Department of Physics, University of California, Santa Barbara, CA 93106, USA 
             }
 
\abstract{
  Multiple modifications of general relativity (GR) have been proposed in the literature in order to understand the nature of the accelerated expansion of the Universe. However, thus far all the predictions of GR have been confirmed with constantly increasing accuracy.
  In this work, we study the imprints of a particular class of models -- ``screened'' modified gravity theories -- on cosmic filaments.
  We have utilized the $N$-body code \texttt{ISIS}/\texttt{RAMSES} to simulate the symmetron model and the Hu-Sawicky $f(R)$ model, and we post-process the output with \texttt{DisPerSE} to identify the filaments of the cosmic web.
  We investigated how the global properties of the filaments -- such as their lengths, masses, and thicknesses -- as well as their radial density and speed profiles change under different gravity theories. We find that filaments are, on average, shorter and denser in modified gravity models compared to in $\Lambda$CDM. We also find that the speed profiles of the filaments are enhanced, consistent with theoretical expectations. 
  Overall, our results suggest that cosmic filaments can be an effective complementary probe of screened modified gravity theories on Mpc scales.
}
   \maketitle
%

\section{Introduction}
The current standard model of cosmology is $\Lambda$CDM.
By introducing two major unknowns in the theory, cold dark matter (CDM) and a cosmological constant ($\Lambda$), this model provides an excellent fit to many observations on large scales, such as the cosmic microwave background \citep{2000ApJ...536L..63M,2002ApJ...571..604N,WMAP9,2016A&A...594A..13P} and the late-time acceleration of the Universe as measured from Type Ia supernovae \citep{1998AJ....116.1009R,1999ApJ...517..565P,2003ApJ...594....1T}. 
There are many alternatives to the $\Lambda$CDM model \citep[see, e.g.,][]{DarkEnergyAT}, and one of the simpler ones is to introduce a scalar field to the Einstein-Hilbert Lagrangian.
When coupled to matter, this scalar field gives rise to an additional gravitational force component -- often called a fifth force \citep[see, e.g.,][]{2006PhRvL..97o1102M,2012PhR...513....1C,2013LRR....16....6A}.
Gravitational tests on Earth and within the Solar System, however, constrain the strength of a potential fifth force to be orders of magnitude lower than its Newtonian counterpart \citep[for a review of local constraints see][]{2006LRR.....9....3W}. This implies that if a scalar field coupled to matter exists in the Universe, the fifth force must be somehow suppressed in certain environments to satisfy the current constraints.
There are many such screening mechanisms that hide the fifth force in high-density regions \citep[see, e.g.,][for reviews]{2010arXiv1011.5909K,2015PhR...568....1J}.

In this paper, we study two different types of screening mechanisms: the symmetron model \citep{2011PhRvD..84j3521H}, and the chameleon screening mechanism \citep{PhysRevD.69.044026}. 
Both models suppress the fifth force in high-density regions but mediate a long range force in low-density regions. However, while the former relates the vacuum expectation value and the coupling strength between the scalar and matter, the latter alters the scalar field mass -- and, thus, the range of the fifth force --- based on the density of the region. In order to detect such a screening mechanism, it is necessary to probe scales beyond the solar system and regions which are not screened. It is therefore worthwhile to search for signatures of these models in the cosmic web of large scale structure. 

The cosmic web is the complex, hierarchical distribution of matter on large scales \citep{1996Natur.380..603B,2018MNRAS.473.1195L}. It is composed of over-dense dark matter halos, in which galaxies reside; long, filamentary structures that connect and feed the halos; flat walls or ``pancakes''; and vast, under-dense voids. The web-like arrangement of dark matter is reflected in the galaxies, which are biased tracers of the underlying density field, as has been observed by, for example, the Sloan Digital Sky Survey \citep[SDSS;][]{2004ApJ...606..702T} and the 2MASS redshift survey \citep{2012ApjS..199...26H}.

Different structures of the cosmic web have previously been explored as potential testbeds for modified gravity. In particular, dark matter halos and clusters have been studied extensively in the literature, including their mass functions \citep{PhysRevD.81.103002,PhysRevD.83.063503,2012JCAP...10..002B,2013JCAP...04..029B,2012ApJ...748...61D,2018arXiv180607400H}, density profiles \citep{PhysRevD.85.102001}, velocity statistics \citep{2014MNRAS.440..833A,Hellwing2014PhRvL,2015MNRAS.449.2837G} and even internal properties, such as their spin \citep{2017MNRAS.468.3174L}. There has been some attention to the effect of screening on the general structures of the cosmic web \citep{2014JCAP...07..058F,2015JCAP...07..049F}, and more recently, cosmic voids have been investigated a potentially powerful probe of modified gravity theories \citep{Cai2015MNRAS,Zivick2015MNRAS,Voivodic2017PhRvD,2018MNRAS.475.3262F,Cautun2018MNRAS}. 

In this paper, we explore the effect of symmetron and chameleon screening mechanisms on the filaments of the cosmic web. The filaments are good candidates to probe modified gravity, as their relatively low mass-density means they are unlikely to be screened. 
Detecting signatures of the fifth force may be easier in filaments compared to halos, where the enhancement is mostly seen near the edge of the halos \citep[see, e.g.,][]{2015JCAP...07..049F}. In spite of this, filaments were previously mostly ignored as a probe of (screened) modified gravity.

$N$-body simulations are essential tools for studying the evolution of the cosmic web due to its non-linear nature \citep[see reviews, e.g., by][]{2005CSci...88.1088B,2011EPJP..126...55D}. Various gravitational $N$-body simulation codes have been developed which solve for the evolution of an additional gravitational force component stemming from different modified gravity theories \citep[see][for a comparison of modified gravity simulation codes]{2015MNRAS.454.4208W}. We will use the $N$-body code \texttt{ISIS} \citep{2014A&A...562A..78L}, which takes into account a fifth force mediated by an additional scalar field and includes both the symmetron \citep{2011PhRvD..84j3521H} and Hu-Sawicky $f(R)$ models \citep{PhysRevD.76.064004}.

The identification of cosmic filaments in $N$-body simulations is a non-trivial task as filaments are harder to define and identify than halos or voids. Filament identification techniques vary greatly and can be based on graph theory, stochastic point processes, or topological methods \citep{sousbie2011persistent,Alpaslan2014MNRAS,TEMPEL201617,2018MNRAS.473.1195L}. For this work, we will use \texttt{DisPerSE} \citep{sousbie2011persistent,Sousbie2011illustrations} which identifies persistent topological features in discretely-sampled distributions.

The paper proceeds as follows. In Sect. \ref{sec:Theory}, we briefly introduce the modified gravity theories analyzed in this paper. In Sect. \ref{sec:Methods}, we describe the $N$-body simulations used and our methods for identifying and defining cosmic filaments in the distribution of dark matter particles. We then examine the filament properties and their radial density and speed profiles in Sect. \ref{sec:Results}, and we investigate how these differ in modified gravity compared to $\Lambda$CDM. Finally, we conclude the paper in Sect. \ref{sec:Conclusion}.

\section{Modified gravity}
\label{sec:Theory}
We investigated extended theories of gravity which include an extra canonical scalar degree of freedom, $\phi$, in the gravity sector \citep{DarkEnergyAT}. These can in general be written in the Einstein frame as:
\begin{align}
S = \int d^4x \sqrt{-g}\left[\frac{R}{2\kappa^2} -\frac{1}{2}\partial_\mu \phi \partial^\mu \phi -V(\phi)\right] + S_{\rm M},
\label{eq:EH_action_scalar}
\end{align}
where $M_\mathrm{Pl}$ and $H$ are the the reduced Planck mass $M_\mathrm{Pl}^{-2} \equiv 8\pi G $, $\kappa^2 = 8\pi G$, $R$ is the Ricci scalar, and $S_{\rm M}$ is the matter action. We note that the matter fields couple to the Jordan frame metric $\tilde g_{\mu\nu}$ which is connected to to the Einstein frame through $\tilde g_{\mu\nu}=A^2(\phi) g_{\mu\nu}$, where $A(\phi)$ is the coupling function. Here and henceforth, we indicate quantities in the Jordan frame with $\tilde{\,}$.

When minimizing this action with respect to $g_{\mu\nu}$, an additional force, referred to as the fifth force, arises. The strength of this force can be quantified as $\gamma \equiv |\mathbf{F}_\mathrm{Fifth}|/|\mathbf{F}_\mathrm{N}|$, where $\mathbf{F}_\mathrm{N}$ is the Newtonian gravitational force. 
In particular, local gravity experiments constrain $\gamma \ll 1$ in the solar system \citep{2006LRR.....9....3W,cassini}. However, on larger scales the current constraints are weaker, allowing $\gamma \sim 1$. 

We consider two gravity models with two different screening mechanisms, both of which suppress the fifth force on solar system scales, therefore avoiding gravity bounds \citep{2010arXiv1011.5909K,2015PhR...568....1J}, while still modifying gravity in lower density environments on cosmological scales. 

\subsection{Symmetron}
In the symmetron model \citep{2011PhRvD..84j3521H}, the coupling strength between the scalar field and matter is proportional to the vacuum expectation value (VEV). In regions of low mass density, the VEV becomes large, increasing the coupling strength. On the other hand, in regions of high mass density, the VEV becomes small and decouples from matter, thus the fifth force is screened in these regions.

The action of the symmetron models is given by Eq.~\eqref{eq:EH_action_scalar} with the symmetric potential,
\begin{align}
V(\phi) &= -\frac{1}{2}\mu^2 \phi^2 + \frac{1}{4}\lambda \phi^4,
\end{align}
and the coupling function,
\begin{align}
A(\phi) &= 1 + \frac{\phi^2}{M^2} ,
\end{align}
where $\mu$ and $M$ are mass scales and $\lambda$ is a dimensionless constant. These choices of $V(\phi)$, $A(\phi)$ imply that the effective potential has a parabolic shape if the local matter density is larger than a critical threshold and a ``sombrero'' shape otherwise. The resulting minima at $\phi = 0$ and offset from it yield a screened and unscreened fifth force, respectively.

As in \citet{2012ApJ...756..166W}, we can rewrite the parameters $(\mu,\,M,\,\lambda)$ to more physical ones. The range of the fifth force in a vacuum can be written as
\begin{align}
L = \frac{1}{\sqrt{2\mu}};
\end{align}
the scale factor at the time of the symmetry breaking is given by
\begin{align}
a_{\mathrm{ssb}} = \frac{\Omega_{m0}\rho_{c0}}{\mu^2 M^2};
\end{align}
and finally, the coupling strength can be written as
\begin{align}
\beta = \frac{\mu M_{\mathrm{Pl}}}{M^2\sqrt{\lambda}}.
\end{align}
In the symmetron model, the maximum enhancement of gravity is given by $\gamma_{\rm max}=2\beta^2\left[1-(a_{\rm ssb}/a)^3\right]$ \citep[e.g.,][]{2015A&A...583A.123G}.

\subsection{Hu-Sawicky $f(R)$ gravity}
$f(R)$ gravity is perhaps the most well-studied modified gravity model. In this model, one adds a function, $f(\tilde R)$, of the Ricci scalar to the Einstein-Hilbert action in the Jordan frame:
\begin{align}
S = \int d^4x \sqrt{-\tilde g}\left[\frac{\tilde R}{2\kappa^2} + f(\tilde R)\right] + S_{\rm M}.
\label{eq:Action_fofr}
\end{align}
In this paper, we consider the Hu-Sawicky model \citep{PhysRevD.76.064004}, which makes use of chameleon screening. In this model, $f(\tilde R)$ is given by 
\begin{align}
f({\tilde R}) = -m^2 \frac{c_1({\tilde R}/m^2)^n}{1 + c_2({\tilde R}/m^2)}.
\end{align}
The parameters $c_1, c_2$, and $n$ are positive constants, and $m^2 = H_0^2/\Omega_{m0}$. By requiring that the model yields late-time accelerated expansion, in the form of an effective cosmological constant, the parameters $c_1$ and $c_2$ may be reduced down to a parameter $f_{R0}$ \citep[see, e.g.][]{2014A&A...562A..78L,2017MNRAS.467.1569A}, given as
\begin{align}
f_{R0} = -n \frac{c_1}{c_2^2}\left(\frac{\Omega_{m0}}{3(\Omega_{m0} + 4\Omega_{\Lambda0})} \right)^{n+1}.
\end{align}
The choice of $f_{R0}$ and $n$ fully specifies our models, and they relate to the range of the fifth force in the cosmological background today as
\begin{align}
  \lambda_\phi \approx 3 \sqrt{\frac{(n+1)}{\Omega_{m0} + 4\Omega_{\Lambda0} }}\sqrt{\frac{|f_{R0}|}{10^{-6}}} \mathrm{ Mpc}/h.
\end{align}
The maximum enhancement of gravity for all the $f(R)$ models is $\gamma_{\rm max}=4/3$ \citep[e.g.,][]{2012PhR...513....1C}.

\section{Methods}
\label{sec:Methods}

\subsection{Simulations}
\label{sec:nbody_code}
We use the $N$-body simulation code \texttt{ISIS} \citep{2014A&A...562A..78L}, a modification of \texttt{RAMSES} \citep{2002A&A...385..337T}. \texttt{ISIS} follows the evolution of a scalar field in the quasi-static limit and changes the gravitational force accordingly. 
Each simulation has $512^3$ dark matter particles in a box with side length 256 $\mathrm{Mpc}\,h^{-1}$.
The parameters used for the background cosmology are $(\Omega_{m0}, \Omega_{\Lambda 0}, H_0) = (0.267, 0.733, 72\, \mathrm{km} \, \mathrm{s}^{-1}\mathrm{Mpc}^{-1})$, and the corresponding dark matter particle mass is $m_{\mathrm{p}} = 9.3\times 10^9 M_\odot\, h^{-1}$. In addition to the $\Lambda$CDM model, we run simulations with three sets of parameters for the $f(R)$ model and four for the symmetron model. The parameters used for these models are listed in Table~\ref{table:MG_params}. Here, we list also the range of the fifth force in vacuum, $\lambda_\phi$, and the maximum enhancement of the fifth force, $\gamma_\mathrm{max}$. All simulations are analyzed at redshift $z=0$.

\begin{table}
  \centering
  \caption{Model parameters of the simulation runs.}
  \begin{tabular}{@{}lrr|cc@{}}
    \hline \hline
    Name & $|f_{R0}|$ & $n$ & $\lambda_\phi$\tablefootmark{a} & $\gamma_{\mathrm{max}}$\tablefootmark{b}\\
     & & & (Mpc $h^{-1}$) \\
    \hline
    fofr4 & $10^{-4}$ & $1$ & $\sim 23.7$ & $4/3$ \\
    fofr5 & $10^{-5}$ & $1$ & $\sim 7.5$  & $4/3$ \\
    fofr6 & $10^{-6}$ & $1$ & $\sim 2.3$ & $4/3$ \\
    \hline
  \end{tabular}\\[1em]

  \begin{tabular}{@{}lrrc|c@{}}
    \hline \hline
    Name & $a_{\rm ssb}$ & $\beta$ & $L$\tablefootmark{a} & $\gamma_\text{max}$\tablefootmark{b} \\
         &          &         & (Mpc $h^{-1}$) & \\
    \hline
    symm\_A & $0.50$ & $1.0$ & $1.0$ & $\sim 1.75$ \\
    symm\_B & $0.33$ & $1.0$ & $1.0$ & $\sim 1.93$ \\
    symm\_C & $0.50$  & $2.0$ & $1.0$ & $\sim 7.00$ \\
    symm\_D & $0.25$ & $1.0$ & $1.0$ & $\sim 1.97$ \\
    \hline
  \end{tabular}
  \tablefoot{
The values to the left of the vertical line are the model input parameters, and values to the right of it are derived physical properties which help to interpret our results.\\
\tablefoottext{a}{Range of the fifth force in vacuum.}
\tablefoottext{b}{Maximum enhancement of gravity.}
}
\label{table:MG_params}
\end{table}

\subsection{Identifying cosmic filaments using \texttt{DisPerSE}}
\texttt{DisPerSE}\footnote{\url{http://www2.iap.fr/users/sousbie/web/html/indexd41d.html}} \citep[Discrete Persistent Structures Extractor;][]{sousbie2011persistent} identifies persistent topological features in discretely-sampled distributions such as $N$-body simulations and galaxy catalogs \citep{Sousbie2011illustrations,Laigle2018MNRAS,Malavasi2017MNRAS}. It is based on discrete Morse theory \citep{forman2002user}, which partitions space into a series of $n$-dimensional domains defined by the gradient of a function (e.g., the density field) and the connectivity of critical points (maxima, minima, and saddle points). Structures are identified as components of the Morse-Smale complex of the density field. \texttt{DisPerSe} defines filaments as one-dimensional lines connecting a maximum (critical point of third order) and a two-saddle (critical point of second order). 

\texttt{DisPerSE} assigns a density to each discrete tracer using the Delaunay tessellation \citep{delanue1934sphere,DTFE2000}, which divides the three-dimensional volume into tetrahedra with vertices at the particle positions. The density at each dark matter particle's location is then given by the volume of its surrounding tetrahedra. In \texttt{DisPerSE}, a random subsample of the $N$-body particles can be input to the Delaunay tessellation for computational convenience while retaining the main features of the filaments ~\citep{Sousbie2011illustrations}; in this paper, we use a subsample of $188^3$ particles.

The finite sampling of the density field by discrete tracers such as dark matter particles induces Poisson noise. In \texttt{DisPerSE}, noisy features are removed by filtering the Morse-Smale complex using persistence theory \citep{edelsbrunner2002topological}, which measures the robustness of pairs of critical points. The relative importance of a given critical pair is assessed by its persistence ratio, defined as the relative density contrast of the critical pair, which can be given a significance value in units of ``$\sigma$'' \citep[see Eq. 4 of][]{sousbie2011persistent}. In this work, we choose a value of 4$\sigma$.
 \citet{sousbie2011persistent} has shown that for a 2$\sigma$ threshold, the probability of a filament to be noise is roughly 5\%; increasing this to a 4$\sigma$ threshold reduces the noise probability down to roughly 0.006\%. To determine which $\sigma$ value to use, we compared the filament distribution for different values of sigmas at different particle subsamples and found that the 4$\sigma$ distribution closely resembled the case using a $256^3$ particle subsample at 5$\sigma$.

\begin{figure}
\centering
\includegraphics[width=\linewidth]{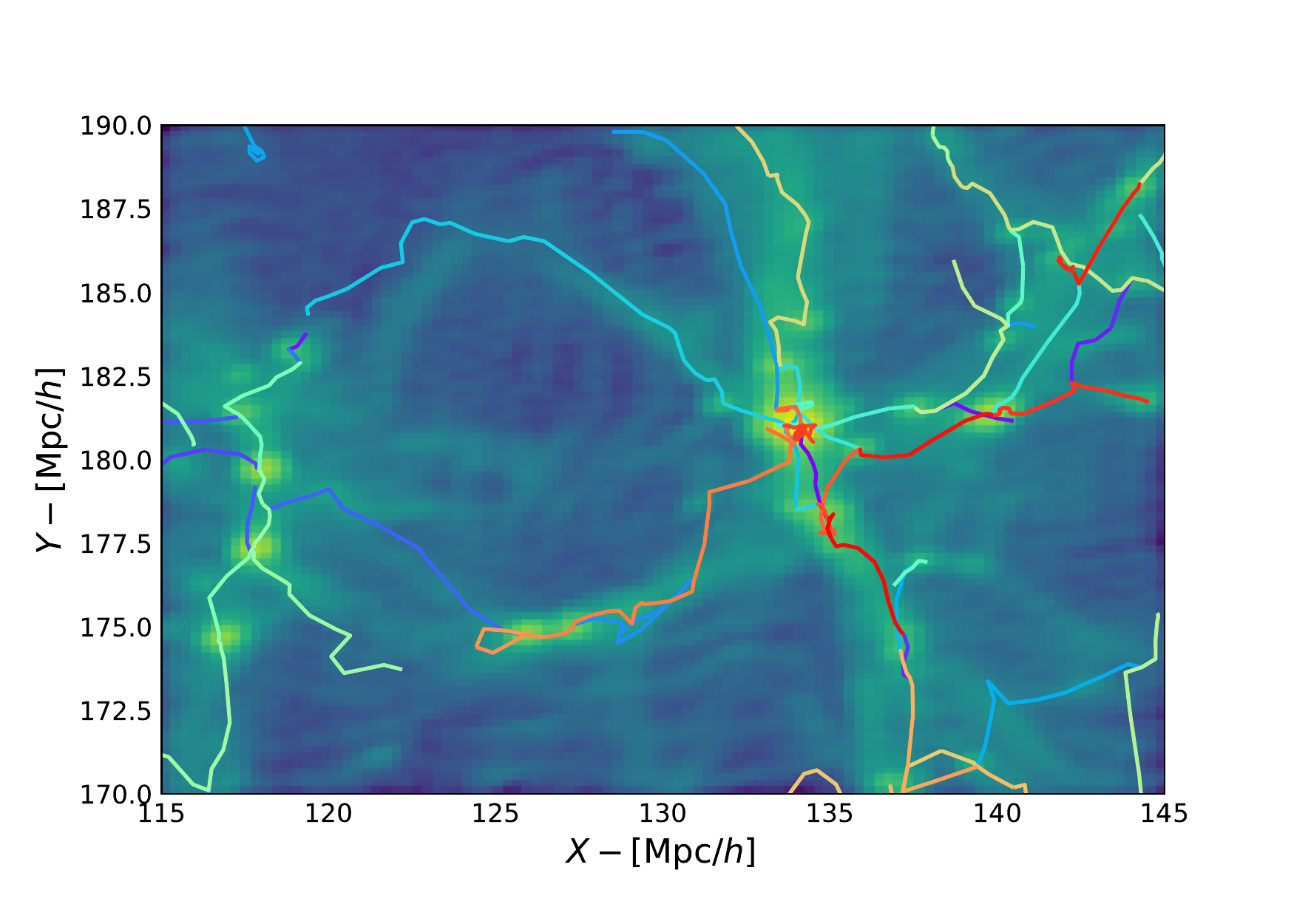}
\caption{An example of the filaments identified by \texttt{DisPerSE}. The plots are taken over a $10.24 \Mpch$ thick slice in the $z$-direction. Each filament in the slice is given a different color, and they are plotted over the background density determined from the dark matter particles.}
\label{fig:Disperse}
\end{figure}

Figure ~\ref{fig:Disperse} illustrates the filaments identified by \texttt{DisPerSE}, which trace the underlying density field of the $N$-body simulation. The endpoints of the filaments connect at clusters, and the endpoint which is defined as the maximum usually resides within a halo identified by AHF \citep{2009ApJS..182..608K}, while most of the two-saddles are found not to be within AHF halos. The output of the \texttt{DisPerSE} code is a list of the coordinates of critical points that define filament segments. In order to measure filament properties, such as mass and density, to study the effect of the fifth force on these filaments, we next need to determine which dark matter particles belong to each filament.

\subsection{Assigning particles to filaments}
\label{sec:DetermineParticles}
The basic steps of our particle-collection procedure are as follows: firstly, for a subset of particles around each filament, calculate their distance to the filament and their distance along the filament; secondly, filter out particles near the end-points of the filaments that are likely to be gravitationally-bound to nearby halos; and finally remove particles further away than the ``edge'' of the filament, determined by its radial density profile.
We note that for this procedure and for the analysis of the results, we use the full set of 512$^3$ particles in each simulation: the sub-sampling was only for the identification of the filament segments. 

\subsubsection{Particle-filament distances}
In order to reduce computing time, we first define a rectangular box around each filament that extends 6\Mpch beyond the maximum extent of the filament, and we only compute distances for particles within this box.
Each filament consists of a set of segments, defined by their endpoints  $\mathbf{s}_1$ and $\mathbf{s}_2$. We want to compute the distance of a particle to the line defined by these endpoints, which can be parameterized as $\mathbf{s}(t) = \mathbf{s}_1 + t(\mathbf{s}_2 - \mathbf{s}_1)$, where $t\in[0,1]$. The minimum distance from this line to a particle located at $\mathbf{p}$ is then $D=|\mathbf{s}(t_s)-\mathbf{p}|$ with
\begin{align}
t_s = \frac{(\mathbf{p} - \mathbf{s_1})\cdot (\mathbf{s_2} - \mathbf{s_1})}{|\mathbf{s_2} - \mathbf{s_1}|^2}.
\end{align}
Since many particles will have a minimum distance to a point on the line outside the bounds given by the endpoints, we enforce that $t_s$ is within the range $[0,1]$ by setting all $t_s < 0$ to $0$ and $t_s > 1$ to $1$.
The final distance to the filament is then the minimum distance for all segments, which we characterize by $t_p=t_s + N$ for the $t_s$ which minimizes $D$, where for the first segment $N=0$, for the second $N=1$, and so on.

\subsubsection{Removing halo particles}
\label{sec:FilterHalos}
As filaments identified by \texttt{DisPerSE}  connect regions of high density, it is very likely that the filaments are contaminated with dark matter particles that are gravitationally bound to halos. Since we are only interested in the particles defining the filament itself, and not the halos, we must find a way to filter out these particles.

We first compute the number density profile along the filament by computing a histogram of the $t_p$ values of all particles in the sub-box around the filament. We then compute the average number of particles in all $t_p$ bins, and we remove the particles belonging to the bins closest to the endpoints of the filament (i.e., the halos) with a number density above this average. If the number density is higher than average in a bin that is not connected to the endpoints. For example, if there is a small density peak within the filament, it is not removed.

\subsubsection{Defining the filament edge}
\label{sec:Filament_thickness_definition}
Once we have removed particles near the endpoints of the filament that may belong to a halo, we next determine the ``edge'' of the filament in order to define its thickness (and also its mass). The edges of structures in $N$-body simulations are notoriously hard to define, even for halos \citep[for a discussion, see][]{Knebe2013MNRAS}.
In the case of halos, the theory of spherical collapse  \citep[e.g.][]{1958RA......5..475L,2002PhR...372....1C} yields a theoretical foundation for a density threshold which divides the collapsed or virialized region from the background. Cosmic filaments, however, are -- as opposed to halos -- non-virialized structures, and a similarly motivated definition does not exist. We thus choose an approach resembling the the commonly-used definition of the virial radius of dark matter halos. The filaments are assumed to consist of segments with cylindrical shape. We then define the thickness as the radius at which the cumulative density profile $\bar\rho(<r)$, that is, the average density within $r$, falls below $10\rho_{\rm c0}$, where $\rho_{\rm c0}$ denotes the critical density today. Filaments that never reach this density threshold are removed from the dataset.

\subsection{Pruning the filament catalog}
At this point we have a catalog of filaments (and their constituent particles), each with a defined length, thickness, and mass. In Section~\ref{sec:Results}, we ignore filaments which fulfill one of the following conditions: 
\begin{enumerate}
\item filaments which are smaller than 1$\mathrm{Mpc} \, h^{-1}$,
\item filaments larger than 20 $\mathrm{Mpc} \, h^{-1}$,
\item filaments with thickness larger than 10 $\mathrm{Mpc} \, h^{-1}$, or
\item filaments with mass larger than $10^{15}M_\odot \, h^{-1}$.
\end{enumerate}
The first point is meant to filter out filaments which are not well resolved due to the mass resolution of our simulation, and removes between 13 to 18\% of all identified filaments in the simulation for the different gravity models. 
The latter three points remove filaments of which there are too few at these sizes (large mass, length, or thickness) to make statistically significant conclusions. We find that less than 1\% of the filaments in our dataset are removed because of the latter three points. The fact that few large filaments are found in our simulations is supported by observations. \citet{2014MNRAS.438.3465T} found that $\lesssim$ 0.01\% of the filament volume, detected in the SDSS galaxy survey, have lengths less than 1 $\mathrm{Mpc}\, h^{-1}$ or larger than  $\sim$30 $\mathrm{Mpc }\, h^{-1}$.

\subsection{Velocity components}
\label{sec:Speed_components}
Though most of our velocity analysis will be focused on the particle speed, that is, the magnitude of its velocity vector, we also decompose the velocity of each particle into radial, $v_{\rm r}$, and tangential, $v_{\rm t}$, components with respect to the segment of the filament that the particle is closest to. The radial component is orthogonal to the filament segment and is defined as positive when the particle is moving away from filament axis. The tangential component runs parallel to the filament segment: we define the positive direction as that which starts from the two-saddle and ends at the maximum point. We note that the third, ``circular'' component is ignored; results for radial and tangential speeds of particles within filaments are given in the Appendix.

\subsection{Error computation}
For the global filament properties, such as the number of filaments at a given length bin, we compute Poissonian errors: $\sigma = \sqrt{N}$, where $N$ is the number of data points in the given bin. For stacked profiles, such as the speed profiles of different filaments, we compute the error on the mean, $\sigma / \sqrt{N}$, where $\sigma$ is the standard deviation of the distribution and $N$ is the number of filaments in the respective stack. Furthermore, the uncertainties for relative differences are computed by Gaussian error propagation.

\section{Results}
\label{sec:Results}
\subsection{Global properties}
In this section, we present the distributions of filament length, mass, and thickness in $\Lambda$CDM and modified gravity simulations, as well as the correlations between these properties. The shaded regions of the global properties show the Poisson noise while the errors for the relative differences are computed by Gaussian error propagation.

\subsubsection{Filament lengths}
\begin{figure}
\centering
\includegraphics[width=\linewidth]{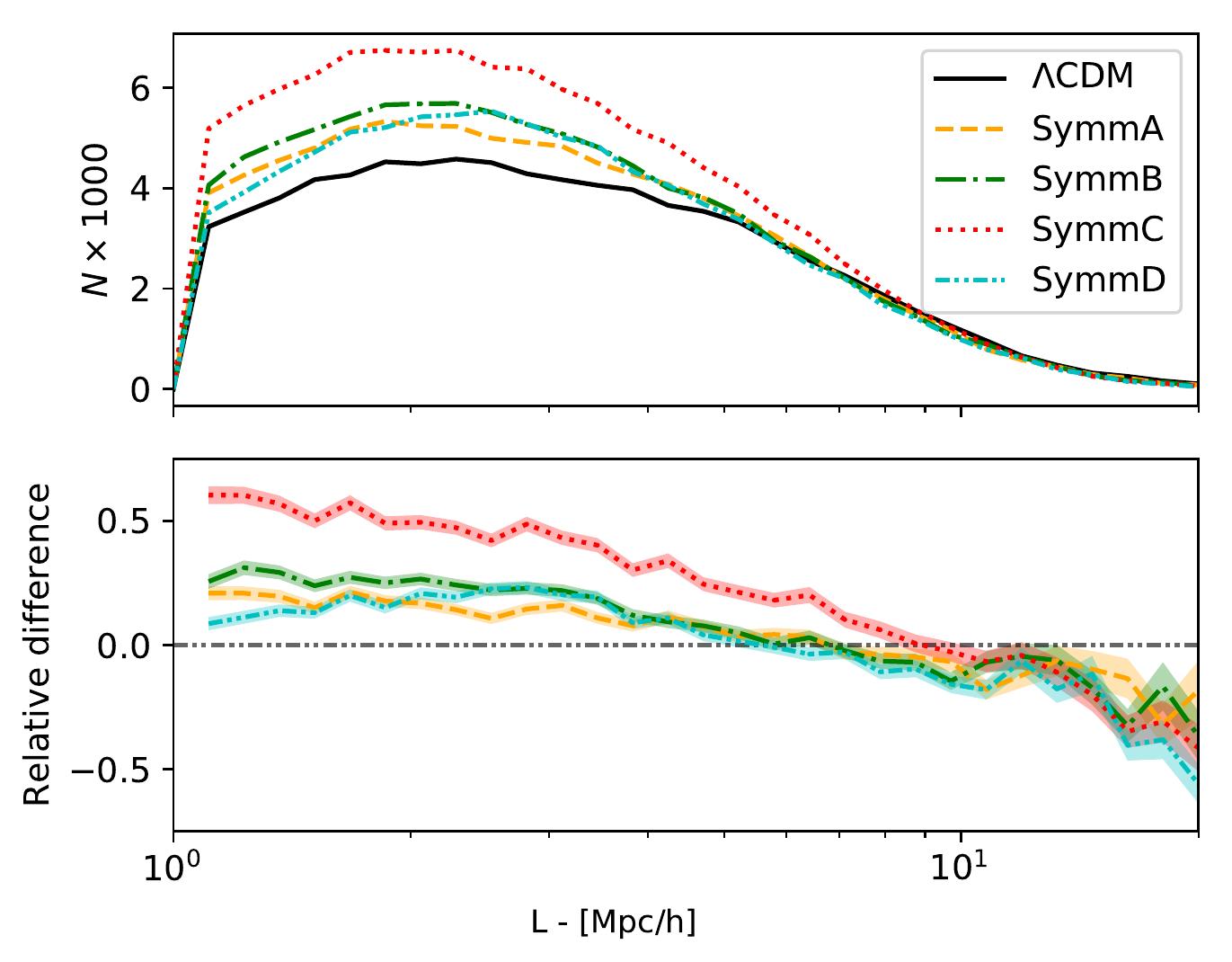}
\caption{Distributions of filament lengths in the $\Lambda$CDM and symmetron models (top), and the relative difference with respect to the $\Lambda$CDM (bottom). The histograms are computed using 20 logarithmically distributed bins.}
\label{fig:Filament_lengths_symmetron}
\end{figure}

\begin{figure}
\centering
\includegraphics[width=\linewidth]{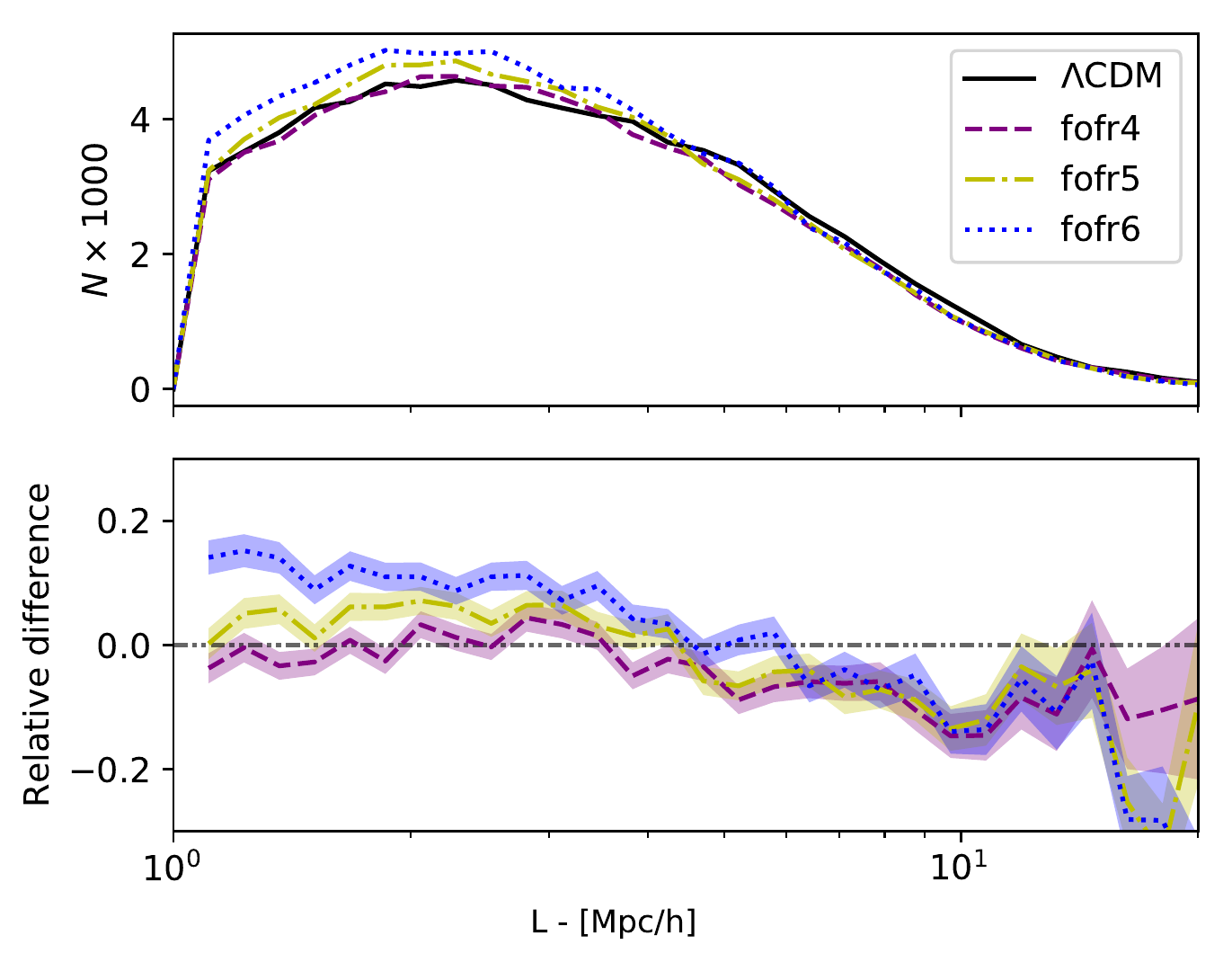}
\caption{Same as Fig. \ref{fig:Filament_lengths_symmetron}, but comparing the $f(R)$ models instead of the symmetron models.}
\label{fig:Filament_lengths_fofr}
\end{figure}

The distributions of the filament lengths in the symmetron and $f(R)$ models are found in Fig. \ref{fig:Filament_lengths_symmetron} and \ref{fig:Filament_lengths_fofr}, respectively. Both figures show the length distributions in the top panel and the relative differences of the modified gravity models with respect to $\Lambda$CDM in the bottom panel. We note that the uncertainties are larger for longer filaments, due to the fact that there are fewer filaments with these lengths.

With the exception of \fofrfour, there are a larger number of short filaments in the modified gravity simulations than in $\Lambda$CDM, where ``short'' is $L\lesssim 10$ Mpc$ \, h^{-1}$ for the symmetron models and $L\lesssim 7$ Mpc $h^{-1}$ for the $f(R)$ models. This is especially true for the \symmC case, which has more than 30\% more filaments at length scales $\sim$1-3 Mpc$ \, h^{-1}$. 
Despite the fact that \fofrsix possesses the shortest fifth force range (cf. \S~\ref{sec:nbody_code}), and is thus closer to $\Lambda$CDM than the other $f(R)$ simulations, it has more short filaments than $\Lambda$CDM, while \fofrfour is the closest to $\Lambda$CDM. While there are, overall, more filaments in each modified gravity model, long filaments appear to be more common in the $\Lambda$CDM case.
The overall shape of the filament length distribution agrees well with the findings of \citet{2014MNRAS.438.3465T} and \citet{2010MNRAS.409..156B}, which used different methods of filament-finding on the SDSS galaxy catalog. 

\subsubsection{Filament masses}
\begin{figure}
\centering
\includegraphics[width=\linewidth]{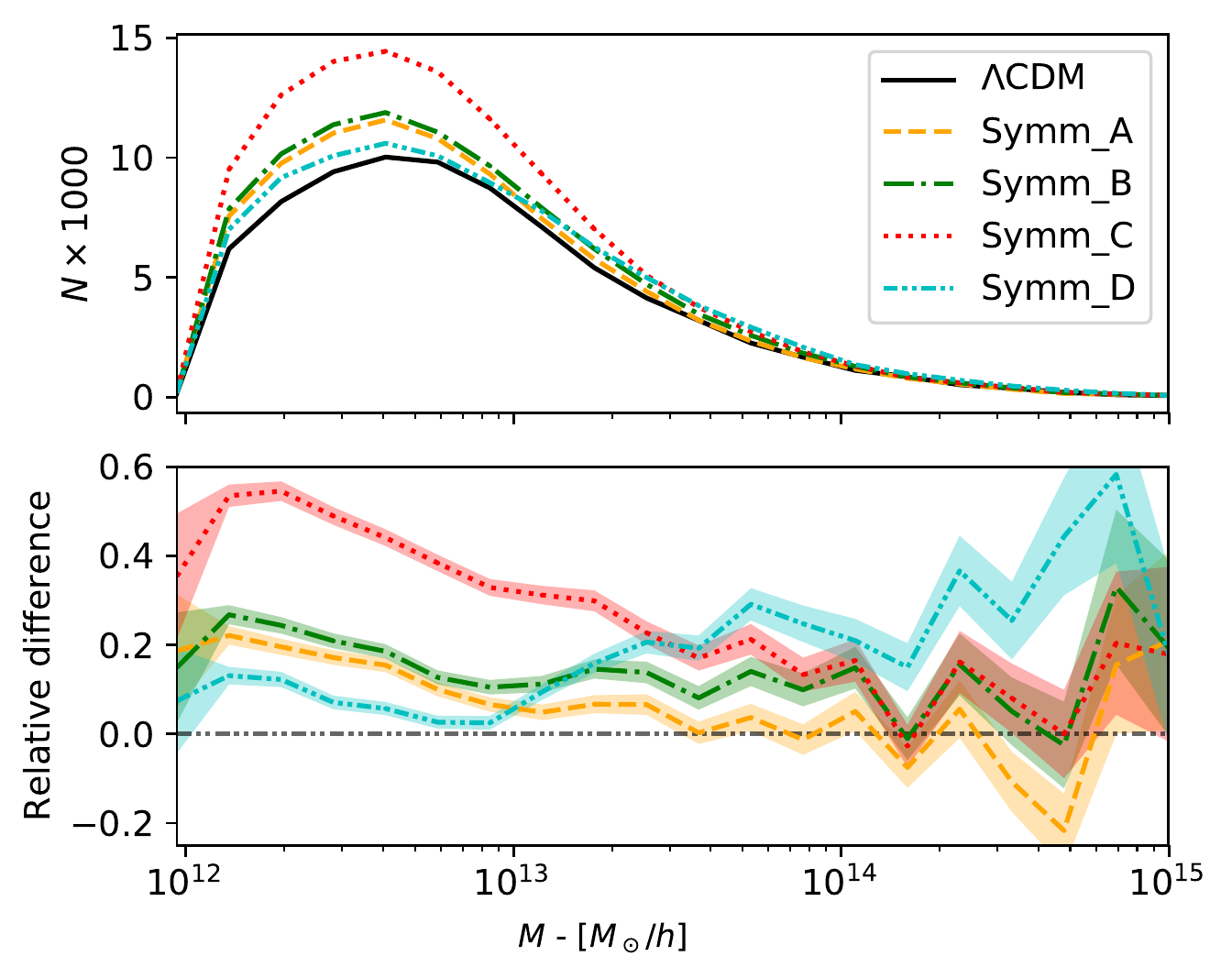}
\caption{Filament mass distributions of the symmetron and $\Lambda$CDM models (top) and the relative difference with respect to $\Lambda$CDM (bottom). The histograms are computed over 20 logarithmically distributed bins.}
\label{fig:Filament_masses_symmetron}
\end{figure}

\begin{figure}
\centering
\includegraphics[width=\linewidth]{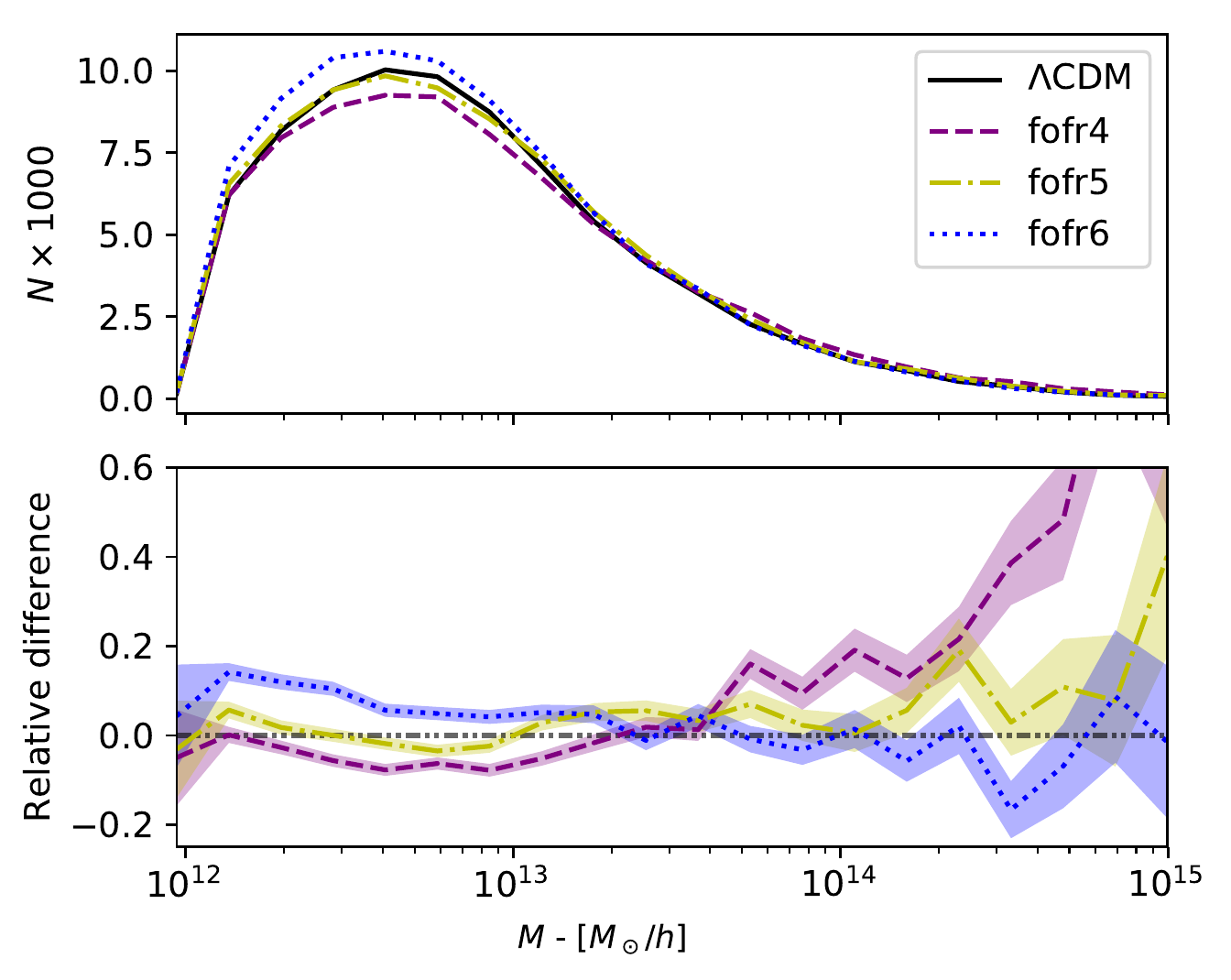}
\caption{Same as Fig. \ref{fig:Filament_masses_symmetron}, but comparing the $f(R)$ models instead of the symmetron models.}
\label{fig:Filament_masses_fofr}
\end{figure}

Figure ~\ref{fig:Filament_masses_symmetron} and \ref{fig:Filament_masses_fofr} show the filament mass distributions of the symmetron and $f(R)$ models, respectively.  
The figures also include, in the bottom panels, the relative differences of the modified gravity models with respect to $\Lambda$CDM. As above, due to the low number of filaments at large masses, the error becomes significantly larger.

The number of filaments, of all mass scales, are generally larger in the symmetron models. Symmetron C has the most filaments of small mass, while \symmD has the most filaments of large mass. In the case of the $f(R)$ models, there are more filaments of small mass in \fofrsix, and more of large mass \fofrfour, compared to in $\Lambda$CDM. A similar result was found for dark matter halos: more of small mass in \fofrsix, and more of large mass in \fofrfour, compared to in $\Lambda$CDM \citep{2013PhRvD..87l3511L, 2011PhRvD..84h4033L}.
We discuss this connection between halos and filaments in more detail in Sec.~\ref{sec:Conclusion}.


\subsubsection{Filament thickness}
\label{sec:thickness}
\begin{figure}
\centering
\includegraphics[width=\linewidth]{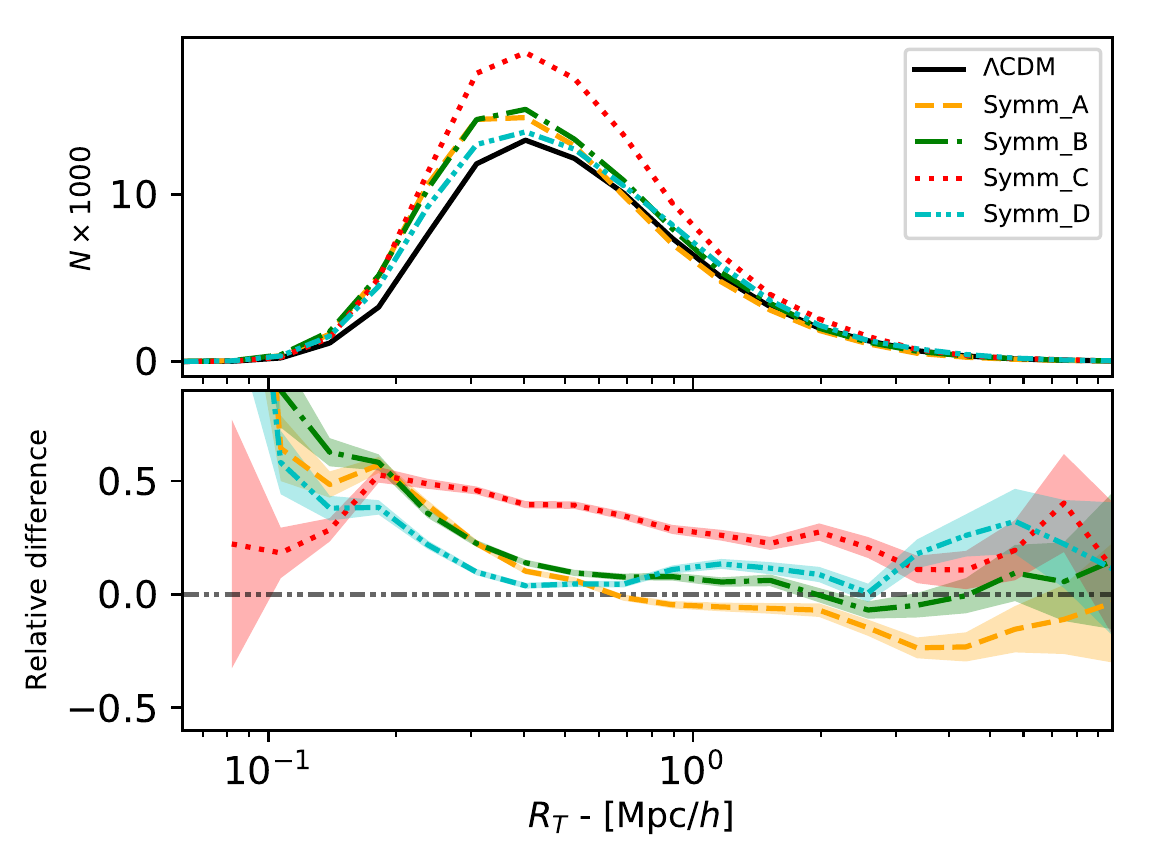}
\caption{Distribution of the filament thicknesses for the $\Lambda$CDM model and the symmetron models (top) and the relative differences with respect to $\Lambda$CDM (bottom). The histograms are computed over 20 logarithmically distributed bins.}
\label{fig:Filament_thickness_symmetron}
\end{figure}

\begin{figure}
\centering
\includegraphics[width=\linewidth]{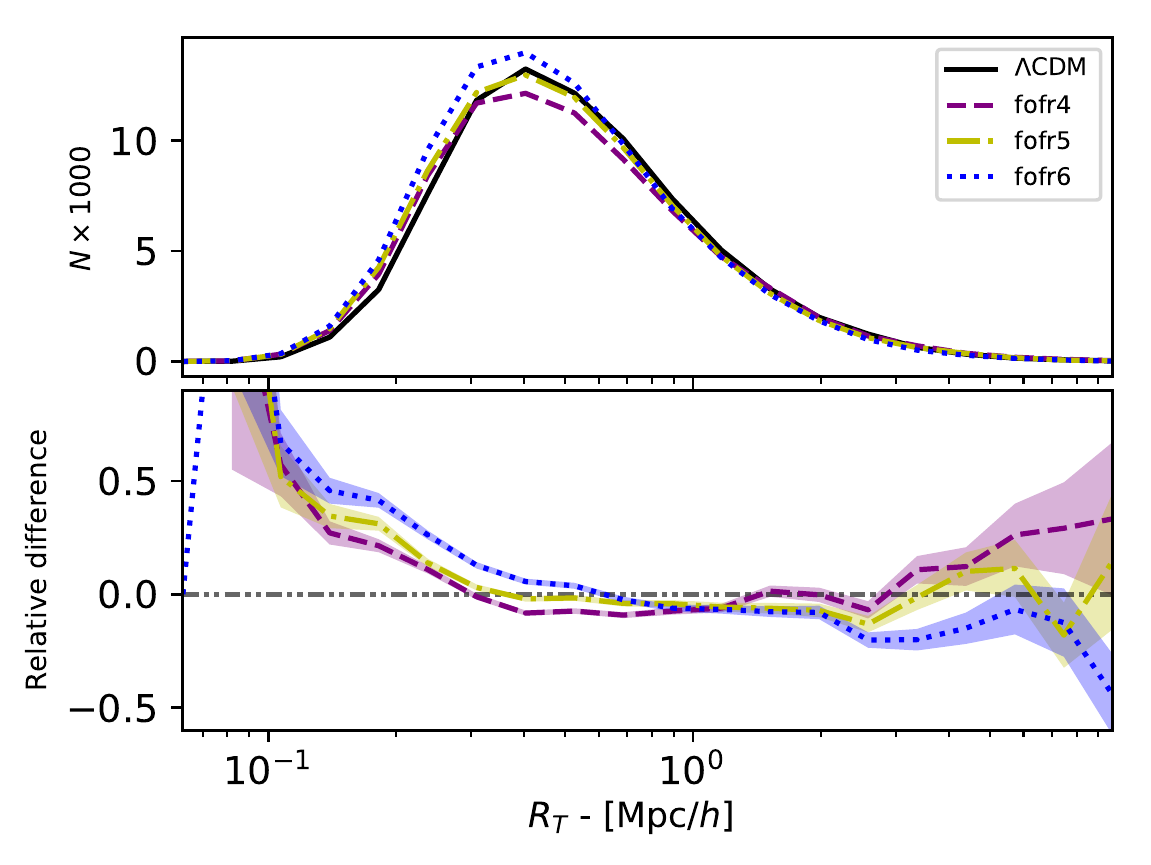}
\caption{Same as Fig. \ref{fig:Filament_thickness_symmetron}, but comparing $f(R)$ models instead of the symmetron models.}
\label{fig:Filament_thickness_fofr}
\end{figure}

In Fig. \ref{fig:Filament_thickness_symmetron}, we show the thickness distributions of the $\Lambda$CDM model and the symmetron models at the top, with the relative differences with respect to $\Lambda$CDM at the bottom. Similar results for the $f(R)$ models are shown in Fig. \ref{fig:Filament_thickness_fofr}. The thickness distribution can be compared to the findings of \citet{2010MNRAS.409..156B}, who
define the filament width based on the shape of the density profile\footnote{Specifically, \citet{2010MNRAS.409..156B} use the root-mean-square of the perpendicular offset of galaxies as width.}, and did not use a density threshold as we did.
As a consequence, their filaments are generally thicker, that is, they
find more filaments with a larger thickness, between $5\Mpch$ to $10\Mpch$.
It is important to note that the thickness of our filaments depend on the density threshold parameter described in Sec. \ref{sec:Filament_thickness_definition}. The overall thickness will increase and decrease if the threshold is set lower and higher, respectively.
While all modified gravity models possess more filaments of thicknesses between $\sim 0.1 \Mpch$ and $0.3 \Mpch$, the differences with respect to $\Lambda$CDM are negligible for filaments with thicknesses between $0.3 \Mpch$ and $2 \Mpch$. The one exception is the \symmC model, which has more than $\sim$20\% filaments in all thickness ranges.
The recovered thickness is naturally linked with the filaments' (orthogonal) density profile which we discuss in \S~\ref{sec:Density}.

\subsection{Correlations between the filament properties}
In this section, we show how the three global properties (mass, length, and thickness) correlate with each other. Intuitively, they should all be correlated. For instance, a long filament, or even a thick filament, should have an increased mass because it covers a larger amount of space and thus includes more particles in the filament. The errors are computed by the standard deviation, while the error of the relative differences are still computed by Gaussian error propagation.

\begin{figure}
\centering
\includegraphics[width=\linewidth]{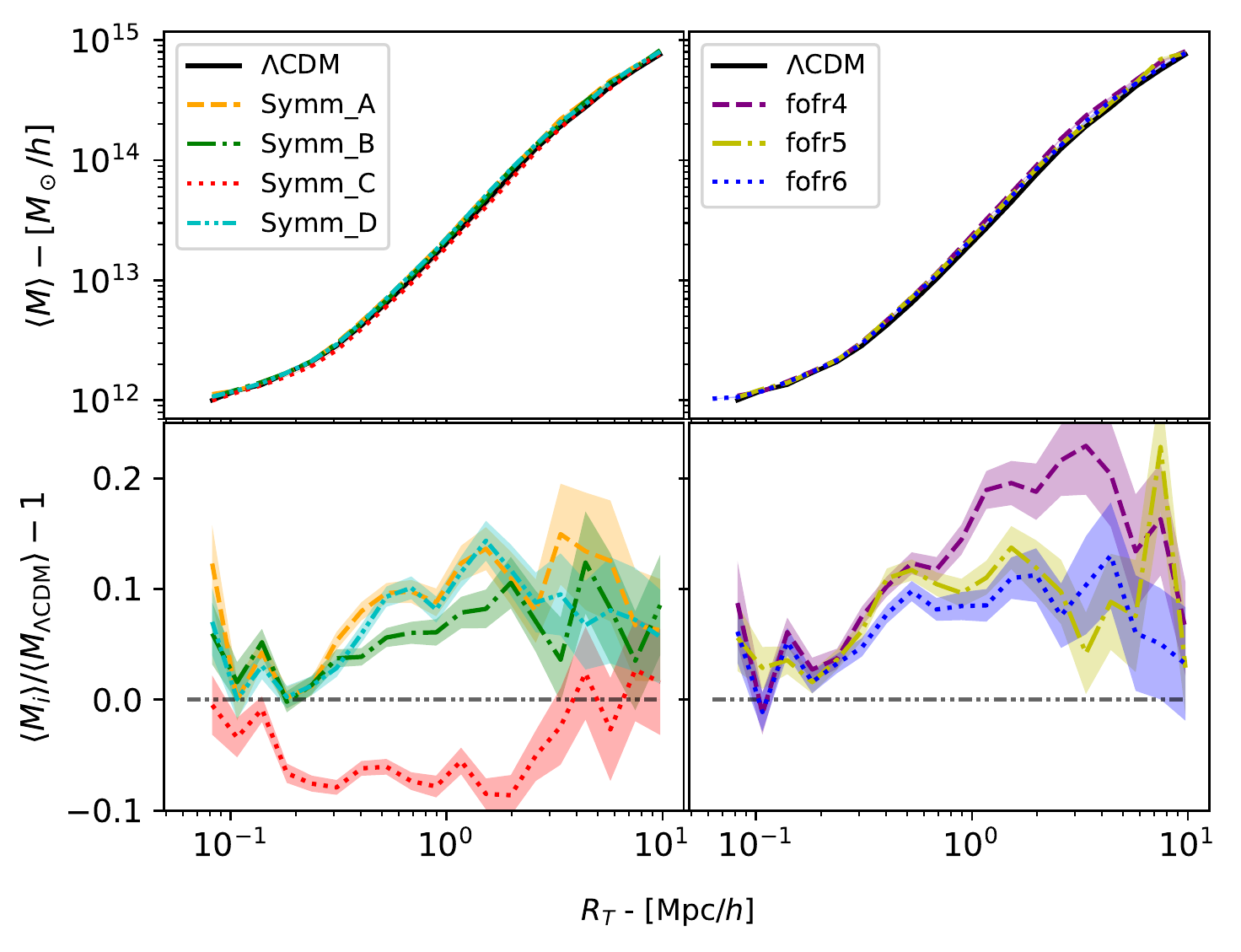}
\caption{\emph{Top}: The average mass of the filaments with a given thickness. \emph{Bottom}: The relative difference of the average mass, with $\Lambda$CDM as the base model.}
\label{fig:MassVSThickess_all}
\end{figure}

The average filament mass as a function of the filament thickness is shown in the top row of Fig.~\ref{fig:MassVSThickess_all}, and the relative differences with respect to the $\Lambda$CDM model are shown in the bottom row. As we can see, the average mass of the filament does indeed increase as the thickness of the filament becomes bigger. This is not surprising, because a thicker filament would contain more particles. This result also agrees with the findings of \citet{2014MNRAS.441.2923C} (see, e.g., their figure~39), which compares the filament linear density with the filament diameter. The authors also conclude that thicker filaments have higher masses.
This correlation is enhanced in all analyzed modified gravity models -- with the exception of the \symmC model.

\begin{figure}
\centering
\includegraphics[width=\linewidth]{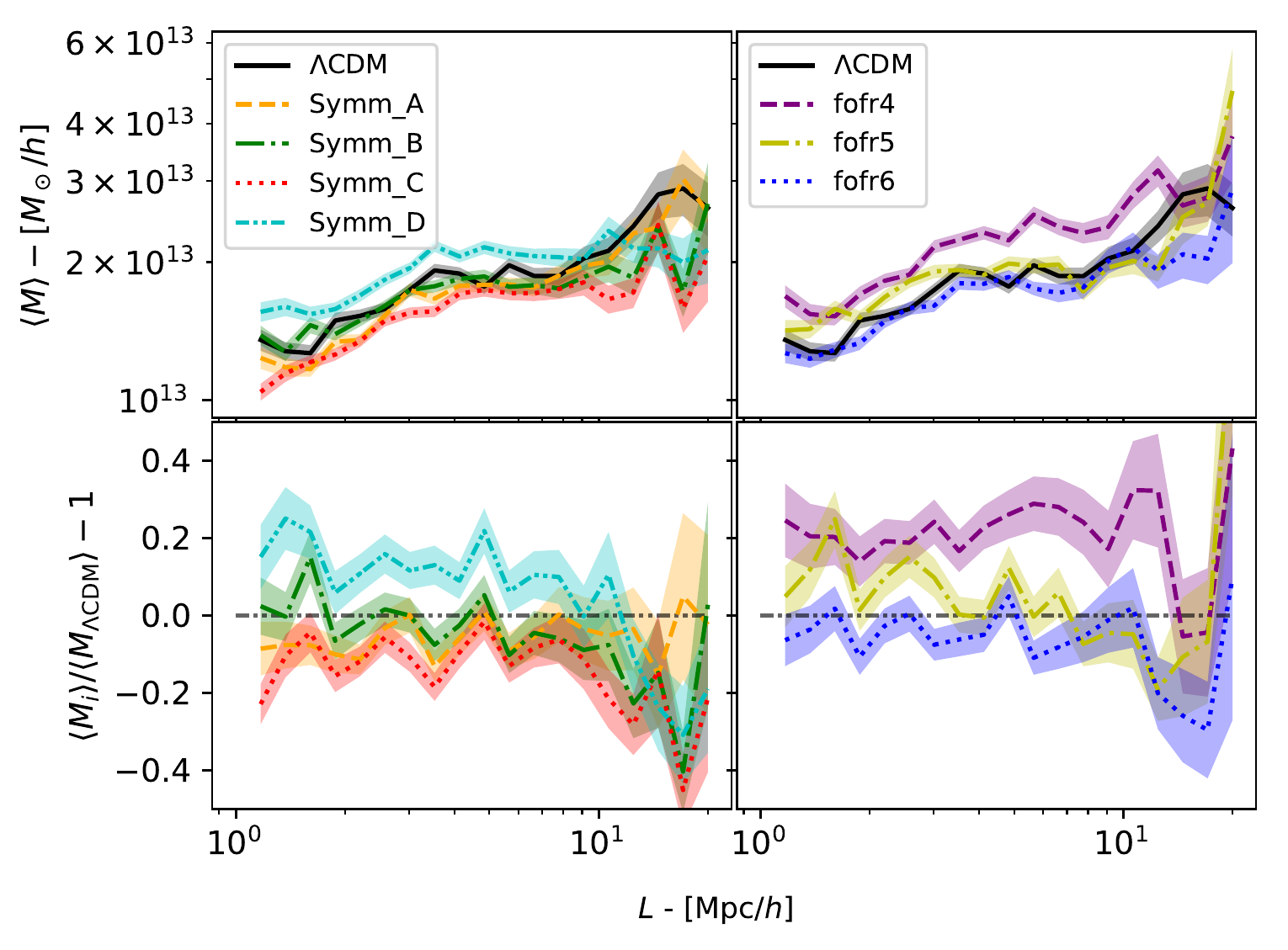}
\caption{\emph{Top}: The average mass of the filaments with a given length. \emph{Bottom}: The relative difference of the modifies gravity models with respect to the $\Lambda$CDM model}
\label{fig:MassVSLength_all}
\end{figure}

The average mass as a function of the filament length is shown in the top row of Fig. \ref{fig:MassVSLength_all}, with the relative differences to $\Lambda$CDM in the bottom row. The average mass does indeed increase as the filament becomes longer, for the same reason as an increased thickness: it contains more particles. 
However, the correlation between mass and length is much weaker than between mass and thickness. The difference between the average masses of the modified gravity theories with respect to $\Lambda$CDM varies in this case. Some models, such as \fofrfour and \symmD, have a larger average mass, while other models, such as \fofrsix and \symmC, have a smaller average mass.

\begin{figure}
\centering
\includegraphics[width=\linewidth]{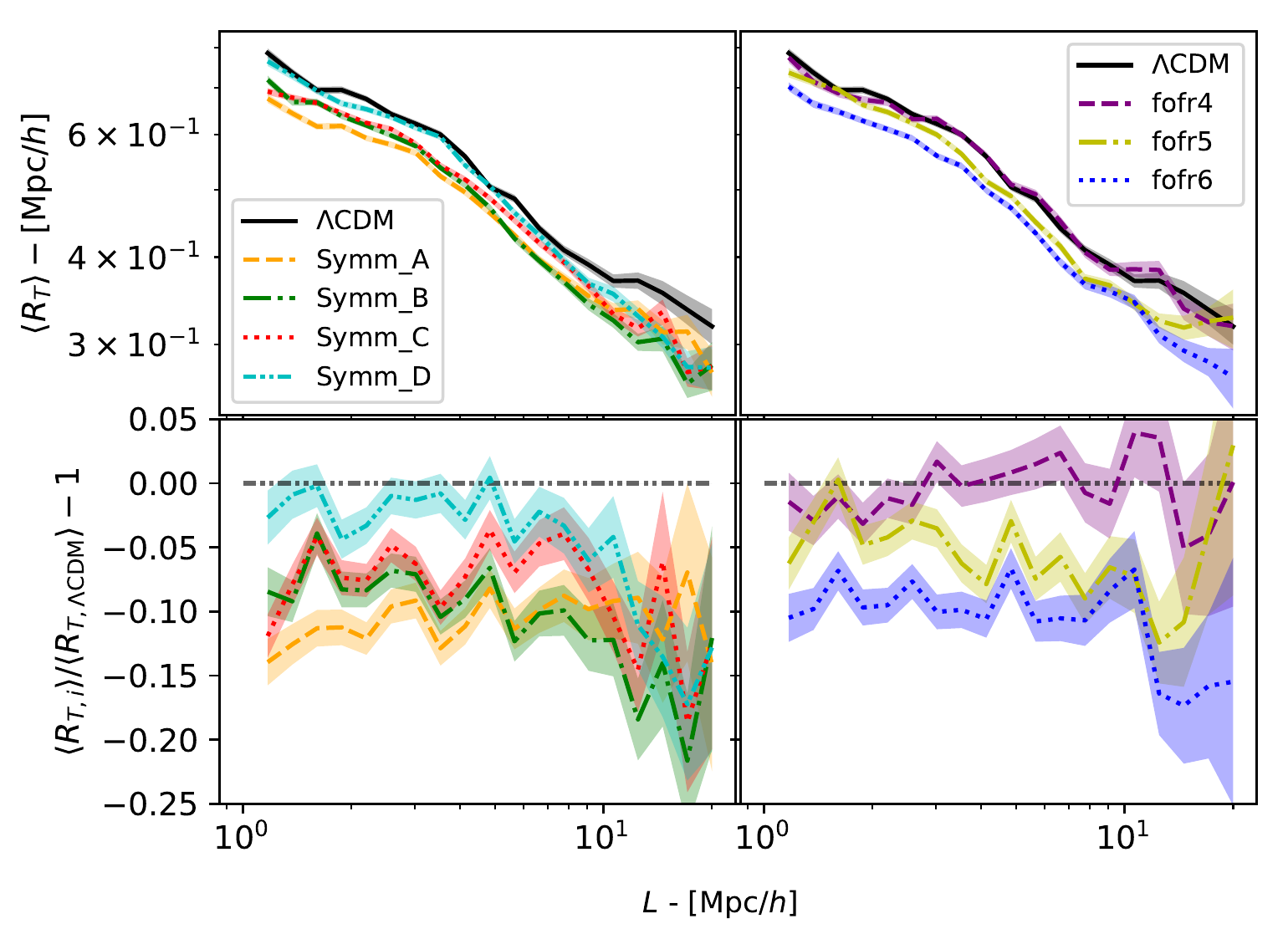}
\caption{\emph{Top}: The average thickness of the filaments with a given length. \emph{Bottom}: Relative differences with respect to the $\Lambda$CDM model.}
\label{fig:ThicknessVSLength_all}
\end{figure}

Finally, we compare the average thickness for filaments of different length in the top row of Fig. \ref{fig:ThicknessVSLength_all}, and the relative differences with respect to $\Lambda$CDM are shown in the bottom row. Interestingly, the thickness decreases as the filament becomes longer. This anticorrelation can be explained due to the fact that, by definition, the filaments start and end at critical points in the density field, and even after potential halo particles are removed (as described in \S~\ref{sec:FilterHalos}), the endpoints of the filaments are still overdense compared to the interior regions. Shorter filaments have these overdense endpoints closer together, resulting in denser, and thus thicker, filaments. The filaments of almost all modified gravity models have, on average, a smaller thickness than in $\Lambda$CDM at all length scales, with \fofrfour being the closest to $\Lambda$CDM. 

\subsection{Density profiles}
\label{sec:Density}
\begin{figure}
\centering
\includegraphics[width=\linewidth]{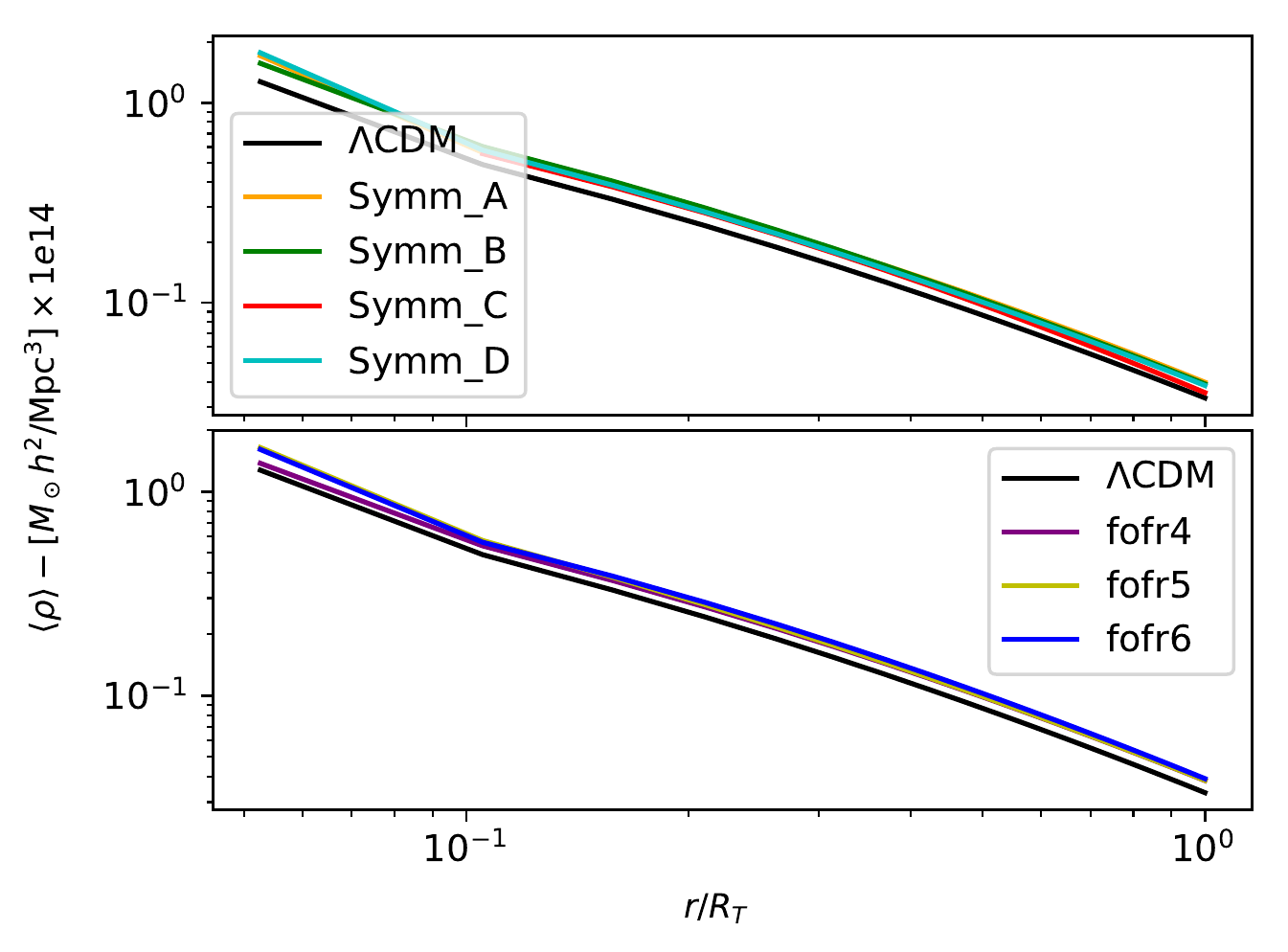}
\caption{The average density profiles of all filaments in each model. On average, the filaments have a higher density near their centers and lower near their outermost radii. The average density is computed as a function of the normalized radius, $r/R_T$.}
\label{fig:Density_plot_all_example}
\end{figure}

We first study the radial density profiles of cosmic filaments and how they change under different gravity models.
We only present the relative difference of the density profiles with respect to the $\Lambda$CDM because the profiles themselves are very similar. 
The average density profiles for all filaments in each model are shown in Fig. \ref{fig:Density_plot_all_example}. This figure shows that, on average, the filaments have a larger density in the center than the outer edge.
The displayed uncertainties of the density profiles are the error of the mean (of the corresponding stack). The uncertainties on the relative differences follow through Gaussian error propagation.

\begin{figure}
\includegraphics[width=\linewidth]{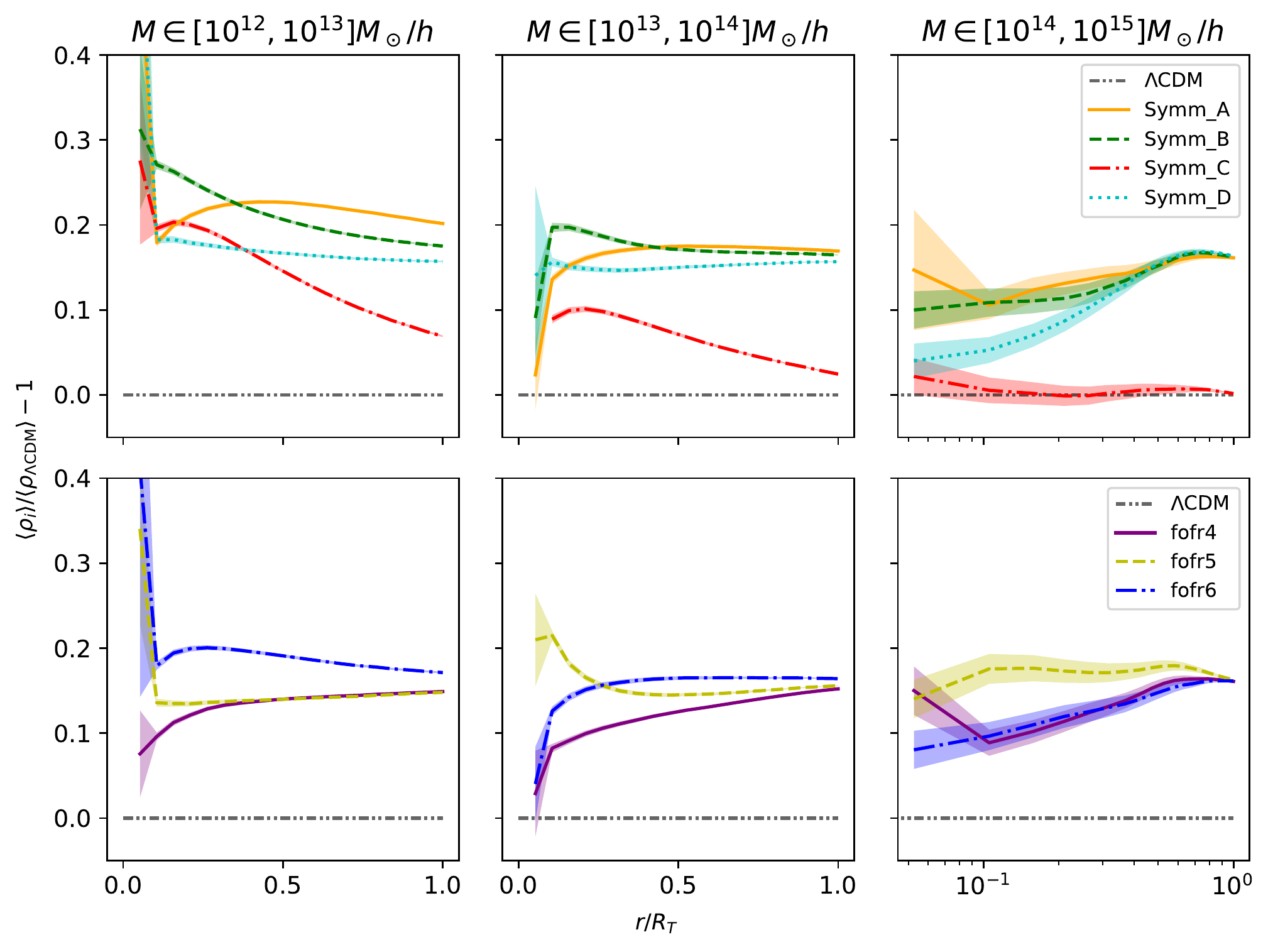}
\caption{The relative difference of the average density profiles of the filaments within a given mass range (shown in their respective titles).}
\label{fig:Density_profile_reldiff_mass}
\end{figure}

To determine whether screening affects filaments of different mass, we consider how the density profiles change at different mass scales. The average density profiles for filaments within the given mass ranges are shown in Fig. \ref{fig:Density_profile_reldiff_mass}. The mass range choice is $M\in[10^{12}, 10^{13}]M_\odot \, h^{-1}$, $M\in[10^{13}, 10^{14}]M_\odot \, h^{-1}$ and $M\in[10^{14}, 10^{15}]M_\odot \, h^{-1}$. 
The average densities are generally larger in the modified gravity models. The difference between \symmC and $\Lambda$CDM becomes smaller as the filament mass becomes larger, and is practically zero for masses $M\geq 10^{14}M_\odot \, h^{-1}$. 
This is most likely due to the screening effect being active for these most massive -- and most dense -- filaments. In the case of \symmD, the density profiles of the most massive filaments deviate more from $\Lambda$CDM in the outer regions of the filament. 
Interestingly, the models that are supposed to deviate the least from $\Lambda$CDM (e.g. \fofrsix\ and \symmA), have, on average, larger densities for filaments of small and intermediate masses.
This implies that cosmic filaments may be a good probe of screened modified gravity models in which halos are already self-screened.

\subsection{Particle velocities and speed profiles}
\label{sec:Speed_profiles}
In this section, we analyze the speeds of particles within filaments, as well as their radial speed profiles. 
Generally, we find that the radial component of the particle velocity accounts for most of the total speed, which means that the tangential speed will only give a small contribution. This also indicates that most particles will move toward the filament as opposed to along it. The results for both the radial and tangential speeds are found in Appendix~\ref{sec:Appendix_rt_speed}. The shaded regions in the plots show the uncertainties computed as for the density profiles.

\begin{figure}
\includegraphics[width=\linewidth]{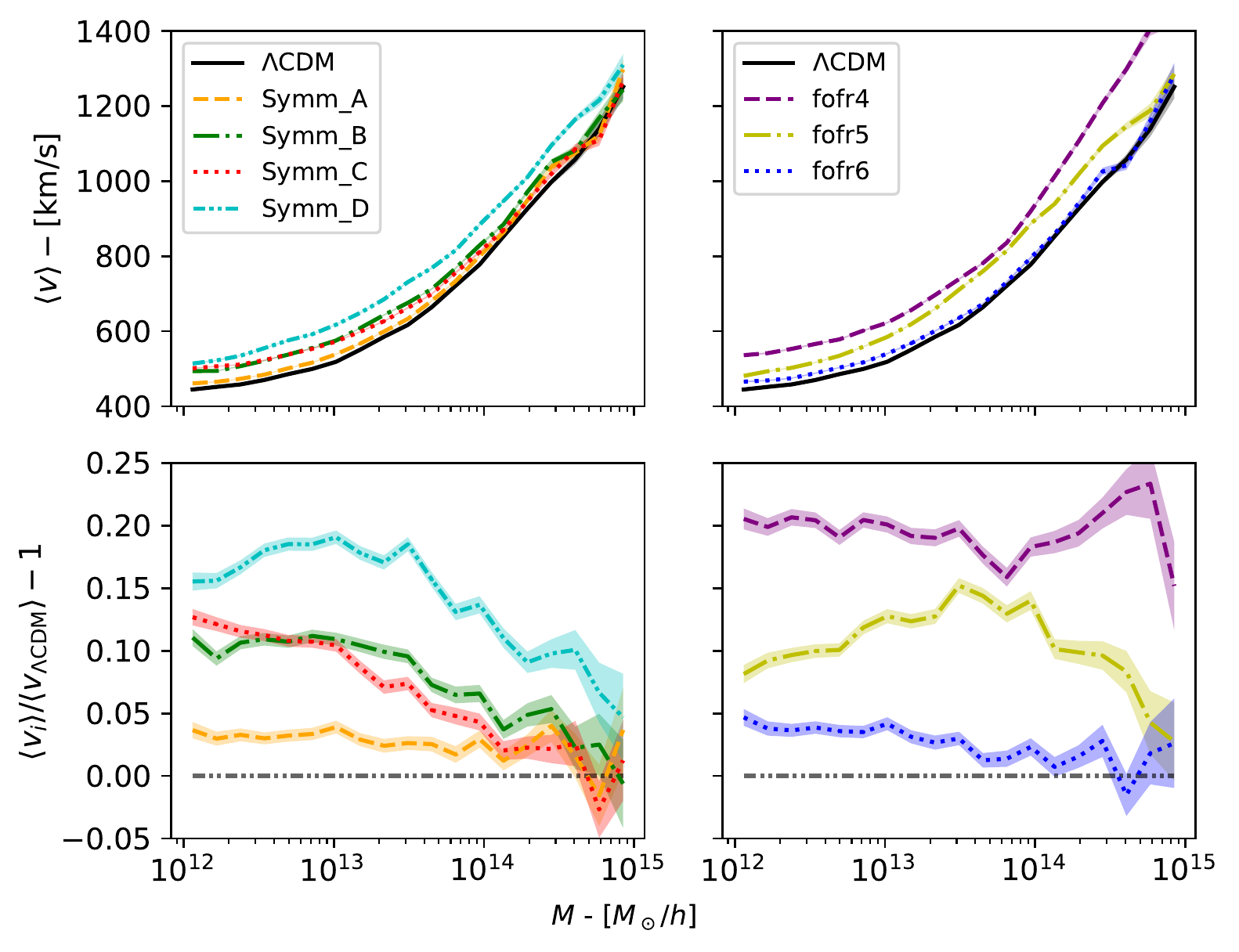}
\caption{The average speed of particles within filaments as a function of filament mass (top), and their relative differences with respect to  $\Lambda$CDM (bottom).}
\label{fig:Average_Speed_mass_bins}
\end{figure}

We show the average speed of the particles within a filament as a function of the filament mass (such that each filament contributes one average particle speed) in the top two panels of Fig. \ref{fig:Average_Speed_mass_bins}, along with the relative differences with respect to $\Lambda$CDM in the bottom two panels.
We find that more massive filaments contain particles with larger speeds, as expected. 
The models  \symmA and \fofrsix deviate the least from $\Lambda$CDM, while \symmD and \fofrfour deviate the most, in their respective classes of models. In general, the difference with respect to $\Lambda$CDM decreases as the filament mass increases, which is likely a result of the screening beginning to be effective for filaments with larger mass.

\begin{figure}
\includegraphics[width=\linewidth]{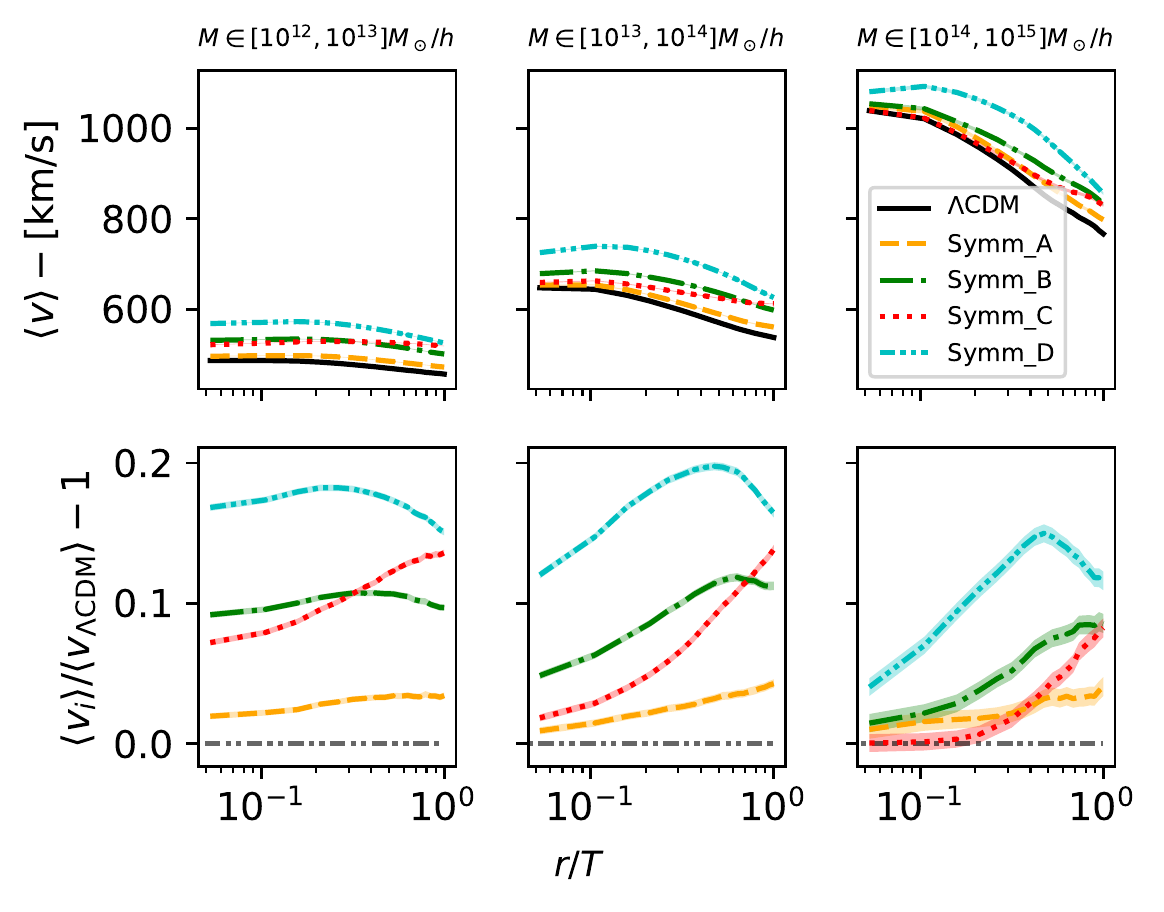}
\caption{The average particle speed (top image) and their relative differences to the $\Lambda$CDM model (bottom image), at different radii, for filaments of the same mass ranges, as given in the respective title. Image compares the $\Lambda$CDM model with the symmetron models. The $x$-axis indicates the distance from the particles divided by the filament thickness, which is taken over 20 bins.}
\label{fig:Average_Speed_mass_symmetron}
\end{figure}

\begin{figure}
\includegraphics[width=\linewidth]{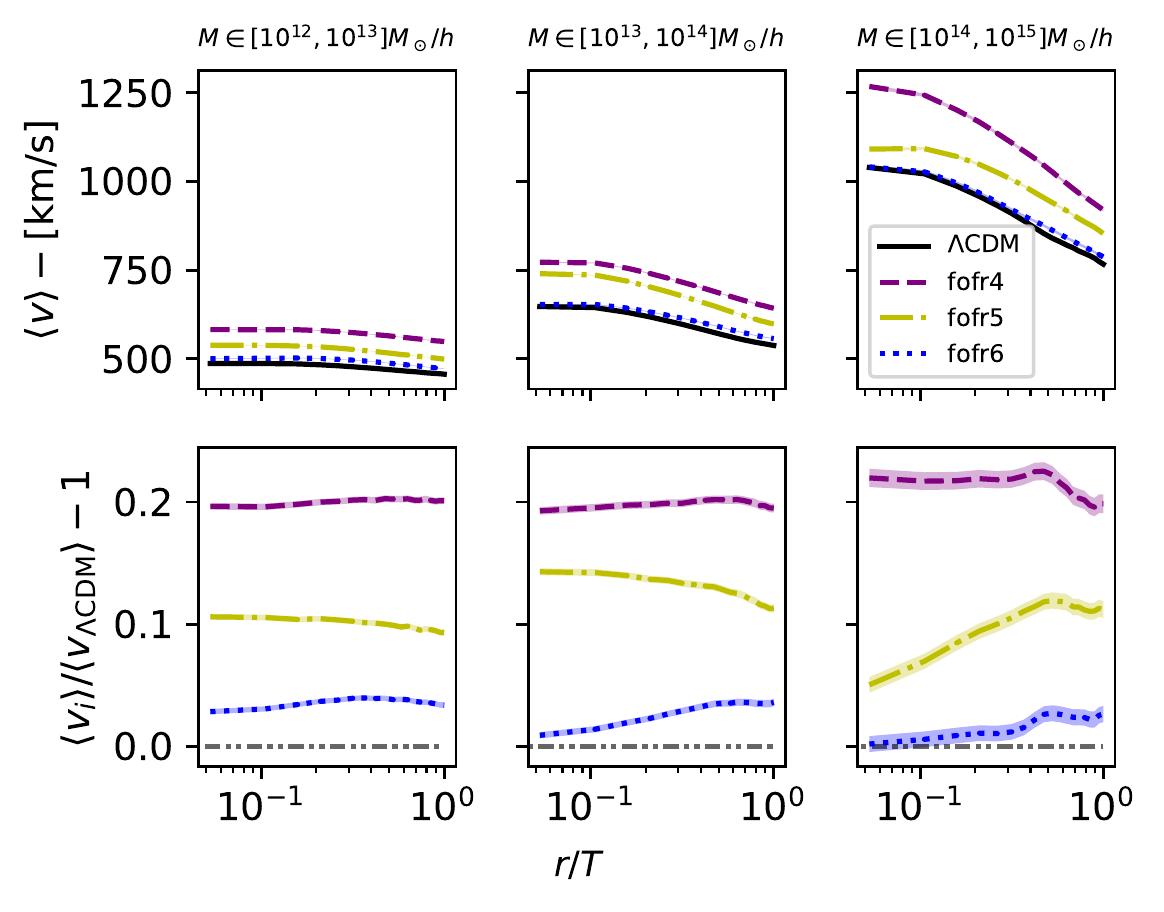}
\caption{Same as Fig. \ref{fig:Average_Speed_mass_symmetron}, but for the $f(R)$ models.}
\label{fig:Average_Speed_mass_fofr}
\end{figure}

The total speed profiles of particles within filaments for filaments of similar mass, for the $\Lambda$CDM \& symmetron and $\Lambda$CDM \& $f(R)$ models, are shown in Fig.~\ref{fig:Average_Speed_mass_symmetron} and \ref{fig:Average_Speed_mass_fofr}, respectively. The top panels show the speed profiles while the bottom images shows the relative difference of the speed profiles with respect to the $\Lambda$CDM model. The mass ranges are $M \in [10^{12}, 10^{13}]M_\odot \, h^{-1}$, $M\in [10^{13}, 10^{14}]M_\odot \, h^{-1}$ and  $M\in [10^{14}, 10^{15}]M_\odot \, h^{-1}$, as they were for the density profiles (see Fig.~\ref{fig:Density_profile_reldiff_mass}). Below, we refer to these mass ranges as filaments of small, intermediate and large masses. 

For all mass ranges, the average speed is generally larger near the filament center, compared to the outer regions of the filament. Similar to speed profiles of clusters of galaxies, this is due to the fact that in the central region a larger fraction of the total energy is kinetic. This also agrees very well with the result of Fig. \ref{fig:Density_plot_all_example}, which indicates that the density is larger, on average, near the center. As expected, by increasing the filament mass, the average speed will also increase because more massive filaments create deeper gravitational potentials. 

The relative difference of the average speed profiles agrees mostly with our theoretical expectations. For instance, comparing the $f(R)$ models with $\Lambda$CDM (see Fig.~\ref{fig:Average_Speed_mass_fofr}), \fofrsix deviates the least from the $\Lambda$CDM, which is expected as the fifth force using this parameter has the smallest range. 
Fig.~\ref{fig:Average_Speed_mass_fofr} also shows that the speed profile of the \fofrsix model follows the typical behavior of screened modified gravity theories: the screening mechanism is active at small radii, leading to a speed profile similar to $\Lambda$CDM. Furthermore, this threshold between screened and unscreened radii shifts inward for more massive filaments, which also happens in dark matter halos \citep{2015JCAP...07..049F}. 
From the same figure, it also becomes clear that filaments are not screened at all in \fofrfour, which causes the deviation with respect to $\Lambda$CDM to become larger in all filament mass ranges.

Figure ~\ref{fig:Average_Speed_mass_symmetron} shows the average speed profiles for the symmetron model. Similarly to the $f(R)$ results, the difference with respect to $\Lambda$CDM is smaller in large filaments and also smaller near the center of the filaments. The differences between the symmetron model and $\Lambda$CDM again coincide with the theoretical expectations. Unsurprisingly, the deviations become larger for larger coupling strengths (such as \symmC compared to \symmA) for the least massive filaments. Nevertheless, the deviation between \symmD and $\Lambda$CDM is the largest for the symmetron models because in this case the screening mechanism becomes active at higher filament densities, compared to the background density. 

\section{Conclusions}
\label{sec:Conclusion}
In this paper, we have explored how filaments of the cosmic web change in different modified gravity models.
First, we have analyzed the global properties of the filaments, such as their masses, lengths, and thicknesses, and studied how they correlate with each other. In addition to the global properties, we have also examined the average density and speed profiles.

We found that the length distribution is overall comparable to the findings of \citet{2014MNRAS.438.3465T} and \citet{2010MNRAS.409..156B}. However, filaments are generally shorter ($L \lesssim 10$ Mpc $h^{-1}$) in all the modified gravity models studied.
While the mass distributions were also comparable to observations \citep{2010MNRAS.409..156B,2014MNRAS.438.3465T}, the screened modified gravity models clearly altered this distribution at the $\gtrsim 10\%$ level. Interestingly, this deviation with respect to $\Lambda$CDM is different for each analyzed model, and can thus potentially be used to constrain the parameter space of screened modified gravity theories. 

Generally, we found that the mass distribution of filaments resembles that previously found for halos \citep[e.g.,][]{PhysRevD.83.063503,2013PhRvD..87l3511L}. For instance, both the halo and filament mass distributions show that \fofrfour and \fofrsix have more filaments and halos of large and small masses, respectively.
This suggests that these deviations in the respective mass functions have similar causes. In screened modified gravity models, halos become (self-)screened if their mass exceeds a certain threshold ($\mu_{200}$ in the notation of \citealp{2015A&A...583A.123G}; also see \citealp{2018MNRAS.477.1133M}). Below this threshold, they grow at an accelerated rate (with the boost given by $\sim \gamma_{\mathrm{max}}$), and above it at the ``normal'' rate set by GR. This leads to a maximum increase in abundance right around this screening threshold where the change in growth rate is largest. 
For instance, amongst the $f(R)$ models analyzed, the fifth force range -- and, consequently, the screening threshold -- is largest for \fofrfour and smallest for \fofrsix (cf. Table~\ref{table:MG_params}). Thus, within this simplistic picture we expect more small mass filaments, in agreement with our findings.

We found  furthermore that the filament thickness distributions did not change dramatically amongst the models analyzed, and the greatest deviations  with respect to $\Lambda$CDM were for filaments with thicknesses 0.1-0.3 Mpc $h^{-1}$. As our definition of the filaments' boundaries is a fixed density threshold, thicker filaments are also denser on average. Thus, the fact that there are no deviations for filaments with thickness $R_{\mathrm{T}}\gtrsim 1\Mpch$ (with the exception of the ``extreme'' \symmC model with $\gamma_{\mathrm{max}}\sim 7$) is a signature of screening.

We also showed that the three global properties are correlated with each other. For instance, thicker filaments are also more massive (cf. \S~4.2). 
However, these correlations change slightly between the different gravity scenarios -- which has to be taken into account when interpreting the deviations of the observables.

Apart from these ``global properties'', we studied the radial density profiles of cosmic filaments, too. 
On average, the density near the center of the filaments is larger than at the edges. As expected, the filaments were denser in the modified gravity models due to the additional fifth force causing the particles to cluster more (with a $\sim 20\%$ increase in matter density).
However, some deviations between modified gravity and $\Lambda$CDM were contrary to the theoretical expectations. 
For instance, in the \fofrsix model, that is, the $f(R)$ model with the shortest fifth force range, the filaments were on average denser than in the \fofrfour case.
Similarly, for the symmetron models, the average density of \symmD, in which the fifth force was active the earliest, was smaller than both \symmA and \symmB.

The velocity field is a direct probe of the underlying gravitational forces and thus a good proxy for the strength of the fifth force.
We found that the filament velocity profiles, for the most part, coincided with our theoretical expectations. For instance, an increased filament mass correlates with a larger average speed, which is due to the fact that a larger filament mass would create a deeper gravitational potential and thus result in a greater kinetic energy. 
Furthermore, the differences in the speed profiles between the modified gravity models and the $\Lambda$CDM model were mostly consistent with the theoretical predictions. For instance, in the symmetron model, \symmA deviates the least from $\Lambda$CDM, while \symmD deviates the most. Interestingly, for the most massive filaments, the fifth force is screened in the \symmC model near the center of the filament.
This can be seen from both the density profiles and the speed profiles.
\\

Overall, we found that the distribution, structure, and dynamics of cosmic filaments are (heavily) altered for some screened modified gravity models. Furthermore, our results show that the probes analyzed were not altered uniformly -- for example, an overall increase for all models -- but instead display a variety of signatures with properties being enhanced, constant, or even decreased with respect to $\Lambda$CDM. While this makes the deviations harder to interpret, it can be used as an advantage, for instance, to distinguish between different modified gravity scenarios. These non-linear changes of the filament properties are due to the fact that the filaments are less dense than halos and thus less likely to be screened. While these results suggest that cosmic filaments can be an excellent probe of modified gravity, the identification and analysis of filamentary structure in emission or absorption data is very challenging, and it remains to be seen whether the resulting data-sets will have much constraining power over the screened modified gravity parameter space.

\begin{acknowledgements}
AH thanks Robert Hagala for discussion as well as numerical assistance during the project. MG thanks Clotilde Laigle for interesting discussions, and the Institute of Theoretical Astrophysics in Oslo for hospitality.
We thank the Research Council of Norway for their support.
The simulations were performed on resources provided by 
UNINETT Sigma2 -- the National Infrastructure for High Performance Computing and 
Data Storage in Norway.
This paper is based upon work from the COST action CA15117 (CANTATA), supported by COST (European Cooperation in Science and Technology). MG acknowledges support from NASA grant NNX17AK58G.
\end{acknowledgements}

\bibliography{MasterProjectReferences}

\begin{thebibliography}{68}
\expandafter\ifx\csname natexlab\endcsname\relax\def\natexlab#1{#1}\fi

\bibitem[{{Alpaslan} {et~al.}(2014){Alpaslan}, {Robotham}, {Driver}, {Norberg},
  {Baldry}, {Bauer}, {Bland-Hawthorn}, {Brown}, {Cluver}, {Colless}, {Foster},
  {Hopkins}, {Van Kampen}, {Kelvin}, {Lara-Lopez}, {Liske}, {Lopez- Sanchez},
  {Loveday}, {McNaught-Roberts}, {Merson}, \& {Pimbblet}}]{Alpaslan2014MNRAS}
{Alpaslan}, M., {Robotham}, A. S.~G., {Driver}, S., {et~al.} 2014, \mnras, 438,
  177

\bibitem[{{Amendola} {et~al.}(2013){Amendola}, {Appleby}, {Bacon}, {Baker},
  {Baldi}, {Bartolo}, {Blanchard}, {Bonvin}, {Borgani}, {Branchini}, {Burrage},
  {Camera}, {Carbone}, {Casarini}, {Cropper}, {de Rham}, {Di Porto}, {Ealet},
  {Ferreira}, {Finelli}, {Garc{\'{\i}}a-Bellido}, {Giannantonio}, {Guzzo},
  {Heavens}, {Heisenberg}, {Heymans}, {Hoekstra}, {Hollenstein}, {Holmes},
  {Horst}, {Jahnke}, {Kitching}, {Koivisto}, {Kunz}, {La Vacca}, {March},
  {Majerotto}, {Markovic}, {Marsh}, {Marulli}, {Massey}, {Mellier}, {Mota},
  {Nunes}, {Percival}, {Pettorino}, {Porciani}, {Quercellini}, {Read},
  {Rinaldi}, {Sapone}, {Scaramella}, {Skordis}, {Simpson}, {Taylor}, {Thomas},
  {Trotta}, {Verde}, {Vernizzi}, {Vollmer}, {Wang}, {Weller}, \&
  {Zlosnik}}]{2013LRR....16....6A}
{Amendola}, L., {Appleby}, S., {Bacon}, D., {et~al.} 2013, Living Reviews in
  Relativity, 16, 6

\bibitem[{Amendola \& Tsujikawa(2010)}]{DarkEnergyAT}
Amendola, L. \& Tsujikawa, S. 2010, Dark Energy (Cambridge University Press)

\bibitem[{{Arnalte-Mur} {et~al.}(2017){Arnalte-Mur}, {Hellwing}, \&
  {Norberg}}]{2017MNRAS.467.1569A}
{Arnalte-Mur}, P., {Hellwing}, W.~A., \& {Norberg}, P. 2017, \mnras, 467, 1569

\bibitem[{{Arnold} {et~al.}(2014){Arnold}, {Puchwein}, \&
  {Springel}}]{2014MNRAS.440..833A}
{Arnold}, C., {Puchwein}, E., \& {Springel}, V. 2014, \mnras, 440, 833

\bibitem[{{Bagla}(2005)}]{2005CSci...88.1088B}
{Bagla}, J.~S. 2005, Current Science, 88, 1088

\bibitem[{{Bennett} {et~al.}(2013){Bennett}, {Larson}, {Weiland}, {Jarosik},
  {Hinshaw}, {Odegard}, {Smith}, {Hill}, {Gold}, {Halpern}, {Komatsu}, {Nolta},
  {Page}, {Spergel}, {Wollack}, {Dunkley}, {Kogut}, {Limon}, {Meyer}, {Tucker},
  \& {Wright}}]{WMAP9}
{Bennett}, C.~L., {Larson}, D., {Weiland}, J.~L., {et~al.} 2013, \apjs, 208, 20

\bibitem[{Bertotti {et~al.}(2003)Bertotti, Iess, \& Tortora}]{cassini}
Bertotti, B., Iess, L., \& Tortora, P. 2003, Nature, 425, 374

\bibitem[{{Bond} {et~al.}(1996){Bond}, {Kofman}, \&
  {Pogosyan}}]{1996Natur.380..603B}
{Bond}, J.~R., {Kofman}, L., \& {Pogosyan}, D. 1996, \nat, 380, 603

\bibitem[{{Bond} {et~al.}(2010){Bond}, {Strauss}, \&
  {Cen}}]{2010MNRAS.409..156B}
{Bond}, N.~A., {Strauss}, M.~A., \& {Cen}, R. 2010, \mnras, 409, 156

\bibitem[{{Brax} {et~al.}(2012){Brax}, {Davis}, {Li}, {Winther}, \&
  {Zhao}}]{2012JCAP...10..002B}
{Brax}, P., {Davis}, A.-C., {Li}, B., {Winther}, H.~A., \& {Zhao}, G.-B. 2012,
  Journal of Cosmology and Astro-Particle Physics, 2012, 002

\bibitem[{{Brax} {et~al.}(2013){Brax}, {Davis}, {Li}, {Winther}, \&
  {Zhao}}]{2013JCAP...04..029B}
{Brax}, P., {Davis}, A.-C., {Li}, B., {Winther}, H.~A., \& {Zhao}, G.-B. 2013,
  Journal of Cosmology and Astro-Particle Physics, 2013, 029

\bibitem[{{Cai} {et~al.}(2015){Cai}, {Padilla}, \& {Li}}]{Cai2015MNRAS}
{Cai}, Y.-C., {Padilla}, N., \& {Li}, B. 2015, \mnras, 451, 1036

\bibitem[{{Cautun} {et~al.}(2018){Cautun}, {Paillas}, {Cai}, {Bose}, {Armijo},
  {Li}, \& {Padilla}}]{Cautun2018MNRAS}
{Cautun}, M., {Paillas}, E., {Cai}, Y.-C., {et~al.} 2018, \mnras, 476, 3195

\bibitem[{{Cautun} {et~al.}(2014){Cautun}, {van de Weygaert}, {Jones}, \&
  {Frenk}}]{2014MNRAS.441.2923C}
{Cautun}, M., {van de Weygaert}, R., {Jones}, B.~J.~T., \& {Frenk}, C.~S. 2014,
  \mnras, 441, 2923

\bibitem[{{Clifton} {et~al.}(2012){Clifton}, {Ferreira}, {Padilla}, \&
  {Skordis}}]{2012PhR...513....1C}
{Clifton}, T., {Ferreira}, P.~G., {Padilla}, A., \& {Skordis}, C. 2012,
  \physrep, 513, 1

\bibitem[{{Cooray} \& {Sheth}(2002)}]{2002PhR...372....1C}
{Cooray}, A. \& {Sheth}, R. 2002, \physrep, 372, 1

\bibitem[{{Davis} {et~al.}(2012){Davis}, {Li}, {Mota}, \&
  {Winther}}]{2012ApJ...748...61D}
{Davis}, A.-C., {Li}, B., {Mota}, D.~F., \& {Winther}, H.~A. 2012, \apj, 748,
  61

\bibitem[{{Dehnen} \& {Read}(2011)}]{2011EPJP..126...55D}
{Dehnen}, W. \& {Read}, J.~I. 2011, European Physical Journal Plus, 126, 55

\bibitem[{Delaunay(1934)}]{delanue1934sphere}
Delaunay, B. 1934, Proceedings of the USSR Academy of Sciences, 793

\bibitem[{Edelsbrunner {et~al.}(2002)Edelsbrunner, Letscher, \&
  Zomorodian}]{edelsbrunner2002topological}
Edelsbrunner, H., Letscher, D., \& Zomorodian, A. 2002, Discrete Comput Geom,
  28, 511

\bibitem[{{Falck} {et~al.}(2015){Falck}, {Koyama}, \&
  {Zhao}}]{2015JCAP...07..049F}
{Falck}, B., {Koyama}, K., \& {Zhao}, G.-B. 2015, Journal of Cosmology and
  Astro-Particle Physics, 2015, 049

\bibitem[{{Falck} {et~al.}(2018){Falck}, {Koyama}, {Zhao}, \&
  {Cautun}}]{2018MNRAS.475.3262F}
{Falck}, B., {Koyama}, K., {Zhao}, G.-B., \& {Cautun}, M. 2018, \mnras, 475,
  3262

\bibitem[{{Falck} {et~al.}(2014){Falck}, {Koyama}, {Zhao}, \&
  {Li}}]{2014JCAP...07..058F}
{Falck}, B., {Koyama}, K., {Zhao}, G.-b., \& {Li}, B. 2014, Journal of
  Cosmology and Astro-Particle Physics, 2014, 058

\bibitem[{Ferraro {et~al.}(2011)Ferraro, Schmidt, \& Hu}]{PhysRevD.83.063503}
Ferraro, S., Schmidt, F., \& Hu, W. 2011, Phys. Rev. D, 83, 063503

\bibitem[{Forman(2002)}]{forman2002user}
Forman, R. 2002, S{\'e}m. Lothar. Combin, 48, 35pp

\bibitem[{{Gronke} {et~al.}(2015{\natexlab{a}}){Gronke}, {Llinares}, {Mota}, \&
  {Winther}}]{2015MNRAS.449.2837G}
{Gronke}, M., {Llinares}, C., {Mota}, D.~F., \& {Winther}, H.~A.
  2015{\natexlab{a}}, \mnras, 449, 2837

\bibitem[{{Gronke} {et~al.}(2015{\natexlab{b}}){Gronke}, {Mota}, \&
  {Winther}}]{2015A&A...583A.123G}
{Gronke}, M., {Mota}, D.~F., \& {Winther}, H.~A. 2015{\natexlab{b}}, \aap, 583,
  A123

\bibitem[{{Hagstotz} {et~al.}(2018){Hagstotz}, {Costanzi}, {Baldi}, \&
  {Weller}}]{2018arXiv180607400H}
{Hagstotz}, S., {Costanzi}, M., {Baldi}, M., \& {Weller}, J. 2018, ArXiv
  e-prints [\eprint[arXiv]{1806.07400}]

\bibitem[{{Hellwing} {et~al.}(2014){Hellwing}, {Barreira}, {Frenk}, {Li}, \&
  {Cole}}]{Hellwing2014PhRvL}
{Hellwing}, W.~A., {Barreira}, A., {Frenk}, C.~S., {Li}, B., \& {Cole}, S.
  2014, \prl, 112, 221102

\bibitem[{{Hinterbichler} {et~al.}(2011){Hinterbichler}, {Khoury}, {Levy}, \&
  {Matas}}]{2011PhRvD..84j3521H}
{Hinterbichler}, K., {Khoury}, J., {Levy}, A., \& {Matas}, A. 2011, \prd, 84,
  103521

\bibitem[{Hu \& Sawicki(2007)}]{PhysRevD.76.064004}
Hu, W. \& Sawicki, I. 2007, Phys. Rev. D, 76, 064004

\bibitem[{{Huchra} {et~al.}(2012){Huchra}, {Macri}, {Masters}, {Jarrett},
  {Berlind}, {Calkins}, {Crook}, {Cutri}, {Erdo{\v g}du}, {Falco}, {George},
  {Hutcheson}, {Lahav}, {Mader}, {Mink}, {Martimbeau}, {Schneider},
  {Skrutskie}, {Tokarz}, \& {Westover}}]{2012ApjS..199...26H}
{Huchra}, J.~P., {Macri}, L.~M., {Masters}, K.~L., {et~al.} 2012, \apjs, 199,
  26

\bibitem[{{Joyce} {et~al.}(2015){Joyce}, {Jain}, {Khoury}, \&
  {Trodden}}]{2015PhR...568....1J}
{Joyce}, A., {Jain}, B., {Khoury}, J., \& {Trodden}, M. 2015, \physrep, 568, 1

\bibitem[{{Khoury}(2010)}]{2010arXiv1011.5909K}
{Khoury}, J. 2010, ArXiv e-prints [\eprint[arXiv]{1011.5909}]

\bibitem[{Khoury \& Weltman(2004)}]{PhysRevD.69.044026}
Khoury, J. \& Weltman, A. 2004, Phys. Rev. D, 69, 044026

\bibitem[{{Knebe} {et~al.}(2013){Knebe}, {Pearce}, {Lux}, {Ascasibar},
  {Behroozi}, {Casado}, {Moran}, {Diemand}, {Dolag}, {Dominguez-Tenreiro},
  {Elahi}, {Falck}, {Gottl{\"o}ber}, {Han}, {Klypin}, {Luki{\'c}},
  {Maciejewski}, {McBride}, {Merch{\'a}n}, {Muldrew}, {Neyrinck}, {Onions},
  {Planelles}, {Potter}, {Quilis}, {Rasera}, {Ricker}, {Roy}, {Ruiz},
  {Sgr{\'o}}, {Springel}, {Stadel}, {Sutter}, {Tweed}, \&
  {Zemp}}]{Knebe2013MNRAS}
{Knebe}, A., {Pearce}, F.~R., {Lux}, H., {et~al.} 2013, \mnras, 435, 1618

\bibitem[{{Knollmann} \& {Knebe}(2009)}]{2009ApJS..182..608K}
{Knollmann}, S.~R. \& {Knebe}, A. 2009, \apjs, 182, 608

\bibitem[{{Laigle} {et~al.}(2018){Laigle}, {Pichon}, {Arnouts}, {McCracken},
  {Dubois}, {Devriendt}, {Slyz}, {Le Borgne}, {Benoit-L{\'e}vy}, {Hwang},
  {Ilbert}, {Kraljic}, {Malavasi}, {Park}, \& {Vibert}}]{Laigle2018MNRAS}
{Laigle}, C., {Pichon}, C., {Arnouts}, S., {et~al.} 2018, \mnras, 474, 5437

\bibitem[{{Lema{\^\i}tre}(1958)}]{1958RA......5..475L}
{Lema{\^\i}tre}, G. 1958, Ricerche Astronomiche, 5, 475

\bibitem[{{L'Huillier} {et~al.}(2017){L'Huillier}, {Winther}, {Mota}, {Park},
  \& {Kim}}]{2017MNRAS.468.3174L}
{L'Huillier}, B., {Winther}, H.~A., {Mota}, D.~F., {Park}, C., \& {Kim}, J.
  2017, \mnras, 468, 3174

\bibitem[{{Li} \& {Hu}(2011)}]{2011PhRvD..84h4033L}
{Li}, Y. \& {Hu}, W. 2011, \prd, 84, 084033

\bibitem[{{Libeskind} {et~al.}(2018){Libeskind}, {van de Weygaert}, {Cautun},
  {Falck}, {Tempel}, {Abel}, {Alpaslan}, {Arag{\'o}n-Calvo}, {Forero-Romero},
  {Gonzalez}, {Gottl{\"o}ber}, {Hahn}, {Hellwing}, {Hoffman}, {Jones},
  {Kitaura}, {Knebe}, {Manti}, {Neyrinck}, {Nuza}, {Padilla}, {Platen},
  {Ramachandra}, {Robotham}, {Saar}, {Shandarin}, {Steinmetz}, {Stoica},
  {Sousbie}, \& {Yepes}}]{2018MNRAS.473.1195L}
{Libeskind}, N.~I., {van de Weygaert}, R., {Cautun}, M., {et~al.} 2018, \mnras,
  473, 1195

\bibitem[{{Llinares} {et~al.}(2014){Llinares}, {Mota}, \&
  {Winther}}]{2014A&A...562A..78L}
{Llinares}, C., {Mota}, D., \& {Winther}, H. 2014, \aap, 562, A78

\bibitem[{{Lombriser} {et~al.}(2013){Lombriser}, {Li}, {Koyama}, \&
  {Zhao}}]{2013PhRvD..87l3511L}
{Lombriser}, L., {Li}, B., {Koyama}, K., \& {Zhao}, G.-B. 2013, \prd, 87,
  123511

\bibitem[{Lombriser {et~al.}(2012)Lombriser, Schmidt, Baldauf, Mandelbaum,
  Seljak, \& Smith}]{PhysRevD.85.102001}
Lombriser, L., Schmidt, F., Baldauf, T., {et~al.} 2012, Phys. Rev. D, 85,
  102001

\bibitem[{{Malavasi} {et~al.}(2017){Malavasi}, {Arnouts}, {Vibert}, {de la
  Torre}, {Moutard}, {Pichon}, {Davidzon}, {Kraljic}, {Bolzonella}, {Guzzo},
  {Garilli}, {Scodeggio}, {Granett}, {Abbas}, {Adami}, {Bottini}, {Cappi},
  {Cucciati}, {Franzetti}, {Fritz}, {Iovino}, {Krywult}, {Le Brun}, {Le
  F{\`e}vre}, {Maccagni}, {Ma{\l}ek}, {Marulli}, {Polletta}, {Pollo}, {Tasca},
  {Tojeiro}, {Vergani}, {Zanichelli}, {Bel}, {Branchini}, {Coupon}, {De Lucia},
  {Dubois}, {Hawken}, {Ilbert}, {Laigle}, {Moscardini}, {Sousbie}, {Treyer}, \&
  {Zamorani}}]{Malavasi2017MNRAS}
{Malavasi}, N., {Arnouts}, S., {Vibert}, D., {et~al.} 2017, \mnras, 465, 3817

\bibitem[{{Melchiorri} {et~al.}(2000){Melchiorri}, {Ade}, {de Bernardis},
  {Bock}, {Borrill}, {Boscaleri}, {Crill}, {De Troia}, {Farese}, {Ferreira},
  {Ganga}, {de Gasperis}, {Giacometti}, {Hristov}, {Jaffe}, {Lange}, {Masi},
  {Mauskopf}, {Miglio}, {Netterfield}, {Pascale}, {Piacentini}, {Romeo},
  {Ruhl}, \& {Vittorio}}]{2000ApJ...536L..63M}
{Melchiorri}, A., {Ade}, P.~A.~R., {de Bernardis}, P., {et~al.} 2000, \apjl,
  536, L63

\bibitem[{{Mitchell} {et~al.}(2018){Mitchell}, {He}, {Arnold}, \&
  {Li}}]{2018MNRAS.477.1133M}
{Mitchell}, M.~A., {He}, J.-h., {Arnold}, C., \& {Li}, B. 2018, \mnras, 477,
  1133

\bibitem[{{Mota} \& {Shaw}(2006)}]{2006PhRvL..97o1102M}
{Mota}, D.~F. \& {Shaw}, D.~J. 2006, Physical Review Letters, 97, 151102

\bibitem[{{Netterfield} {et~al.}(2002){Netterfield}, {Ade}, {Bock}, {Bond},
  {Borrill}, {Boscaleri}, {Coble}, {Contaldi}, {Crill}, {de Bernardis},
  {Farese}, {Ganga}, {Giacometti}, {Hivon}, {Hristov}, {Iacoangeli}, {Jaffe},
  {Jones}, {Lange}, {Martinis}, {Masi}, {Mason}, {Mauskopf}, {Melchiorri},
  {Montroy}, {Pascale}, {Piacentini}, {Pogosyan}, {Pongetti}, {Prunet},
  {Romeo}, {Ruhl}, \& {Scaramuzzi}}]{2002ApJ...571..604N}
{Netterfield}, C.~B., {Ade}, P.~A.~R., {Bock}, J.~J., {et~al.} 2002, \apj, 571,
  604

\bibitem[{{Perlmutter} {et~al.}(1999){Perlmutter}, {Aldering}, {Goldhaber},
  {Knop}, {Nugent}, {Castro}, {Deustua}, {Fabbro}, {Goobar}, {Groom}, {Hook},
  {Kim}, {Kim}, {Lee}, {Nunes}, {Pain}, {Pennypacker}, {Quimby}, {Lidman},
  {Ellis}, {Irwin}, {McMahon}, {Ruiz-Lapuente}, {Walton}, {Schaefer}, {Boyle},
  {Filippenko}, {Matheson}, {Fruchter}, {Panagia}, {Newberg}, {Couch}, \&
  {Project}}]{1999ApJ...517..565P}
{Perlmutter}, S., {Aldering}, G., {Goldhaber}, G., {et~al.} 1999, \apj, 517,
  565

\bibitem[{{Planck Collaboration} {et~al.}(2016){Planck Collaboration}, {Ade},
  {Aghanim}, {Arnaud}, {Ashdown}, {Aumont}, {Baccigalupi}, {Banday},
  {Barreiro}, {Bartlett}, \& et~al.}]{2016A&A...594A..13P}
{Planck Collaboration}, {Ade}, P.~A.~R., {Aghanim}, N., {et~al.} 2016, \aap,
  594, A13

\bibitem[{{Riess} {et~al.}(1998){Riess}, {Filippenko}, {Challis},
  {Clocchiatti}, {Diercks}, {Garnavich}, {Gilliland}, {Hogan}, {Jha},
  {Kirshner}, {Leibundgut}, {Phillips}, {Reiss}, {Schmidt}, {Schommer},
  {Smith}, {Spyromilio}, {Stubbs}, {Suntzeff}, \&
  {Tonry}}]{1998AJ....116.1009R}
{Riess}, A.~G., {Filippenko}, A.~V., {Challis}, P., {et~al.} 1998, \aj, 116,
  1009

\bibitem[{{Schaap} \& {van de Weygaert}(2000)}]{DTFE2000}
{Schaap}, W.~E. \& {van de Weygaert}, R. 2000, \aap, 363, L29

\bibitem[{Schmidt(2010)}]{PhysRevD.81.103002}
Schmidt, F. 2010, Phys. Rev. D, 81, 103002

\bibitem[{{Sousbie}(2011)}]{sousbie2011persistent}
{Sousbie}, T. 2011, \mnras, 414, 350

\bibitem[{Sousbie {et~al.}(2011)Sousbie, Pichon, \&
  Kawahara}]{Sousbie2011illustrations}
Sousbie, T., Pichon, C., \& Kawahara, H. 2011, \mnras, 414, 384

\bibitem[{{Tegmark} {et~al.}(2004){Tegmark}, {Blanton}, {Strauss}, {Hoyle},
  {Schlegel}, {Scoccimarro}, {Vogeley}, {Weinberg}, {Zehavi}, {Berlind},
  {Budavari}, {Connolly}, {Eisenstein}, {Finkbeiner}, {Frieman}, {Gunn},
  {Hamilton}, {Hui}, {Jain}, {Johnston}, {Kent}, {Lin}, {Nakajima}, {Nichol},
  {Ostriker}, {Pope}, {Scranton}, {Seljak}, {Sheth}, {Stebbins}, {Szalay},
  {Szapudi}, {Verde}, {Xu}, {Annis}, {Bahcall}, {Brinkmann}, {Burles},
  {Castander}, {Csabai}, {Loveday}, {Doi}, {Fukugita}, {Gott}, {Hennessy},
  {Hogg}, {Ivezi{\'c}}, {Knapp}, {Lamb}, {Lee}, {Lupton}, {McKay}, {Kunszt},
  {Munn}, {O'Connell}, {Peoples}, {Pier}, {Richmond}, {Rockosi}, {Schneider},
  {Stoughton}, {Tucker}, {Vanden Berk}, {Yanny}, {York}, \& {SDSS
  Collaboration}}]{2004ApJ...606..702T}
{Tegmark}, M., {Blanton}, M.~R., {Strauss}, M.~A., {et~al.} 2004, \apj, 606,
  702

\bibitem[{Tempel {et~al.}(2016)Tempel, Stoica, Kipper, \& Saar}]{TEMPEL201617}
Tempel, E., Stoica, R., Kipper, R., \& Saar, E. 2016, Astronomy and Computing,
  16, 17

\bibitem[{{Tempel} {et~al.}(2014){Tempel}, {Stoica}, {Mart{\'{\i}}nez},
  {Liivam{\"a}gi}, {Castellan}, \& {Saar}}]{2014MNRAS.438.3465T}
{Tempel}, E., {Stoica}, R.~S., {Mart{\'{\i}}nez}, V.~J., {et~al.} 2014, \mnras,
  438, 3465

\bibitem[{{Teyssier}(2002)}]{2002A&A...385..337T}
{Teyssier}, R. 2002, \aap, 385, 337

\bibitem[{{Tonry} {et~al.}(2003){Tonry}, {Schmidt}, {Barris}, {Candia},
  {Challis}, {Clocchiatti}, {Coil}, {Filippenko}, {Garnavich}, {Hogan},
  {Holland}, {Jha}, {Kirshner}, {Krisciunas}, {Leibundgut}, {Li}, {Matheson},
  {Phillips}, {Riess}, {Schommer}, {Smith}, {Sollerman}, {Spyromilio},
  {Stubbs}, \& {Suntzeff}}]{2003ApJ...594....1T}
{Tonry}, J.~L., {Schmidt}, B.~P., {Barris}, B., {et~al.} 2003, \apj, 594, 1

\bibitem[{{Voivodic} {et~al.}(2017){Voivodic}, {Lima}, {Llinares}, \&
  {Mota}}]{Voivodic2017PhRvD}
{Voivodic}, R., {Lima}, M., {Llinares}, C., \& {Mota}, D.~F. 2017, \prd, 95,
  024018

\bibitem[{{Will}(2006)}]{2006LRR.....9....3W}
{Will}, C.~M. 2006, Living Reviews in Relativity, 9, 3

\bibitem[{{Winther} {et~al.}(2012){Winther}, {Mota}, \&
  {Li}}]{2012ApJ...756..166W}
{Winther}, H.~A., {Mota}, D.~F., \& {Li}, B. 2012, \apj, 756, 166

\bibitem[{{Winther} {et~al.}(2015){Winther}, {Schmidt}, {Barreira}, {Arnold},
  {Bose}, {Llinares}, {Baldi}, {Falck}, {Hellwing}, {Koyama}, {Li}, {Mota},
  {Puchwein}, {Smith}, \& {Zhao}}]{2015MNRAS.454.4208W}
{Winther}, H.~A., {Schmidt}, F., {Barreira}, A., {et~al.} 2015, \mnras, 454,
  4208

\bibitem[{{Zivick} {et~al.}(2015){Zivick}, {Sutter}, {Wandelt}, {Li}, \&
  {Lam}}]{Zivick2015MNRAS}
{Zivick}, P., {Sutter}, P.~M., {Wandelt}, B.~D., {Li}, B., \& {Lam}, T.~Y.
  2015, \mnras, 451, 4215

\end{thebibliography}

\begin{appendix}
\section{Radial and tangential speed components}
\label{sec:Appendix_rt_speed}

We have in \S~\ref{sec:Speed_profiles} only shown the total speed of the particles within the filament. The total speed, however, does not describe the general direction of the particles themselves, such as whether the particles is moving away or toward the filament center or how the particles moves along the filament axis. By intuition, the particles themselves should mostly move toward the filament center. In addition to this, the average tangential component should be close to zero as the particles should not have any preferential direction along the filament axis. Although, this is not true for the particles near the end points, where the high density regions usually reside, but because they move in opposite directions, the average tangential speed should therefore cancel each other out. 

\begin{figure}
\includegraphics[width=\linewidth]{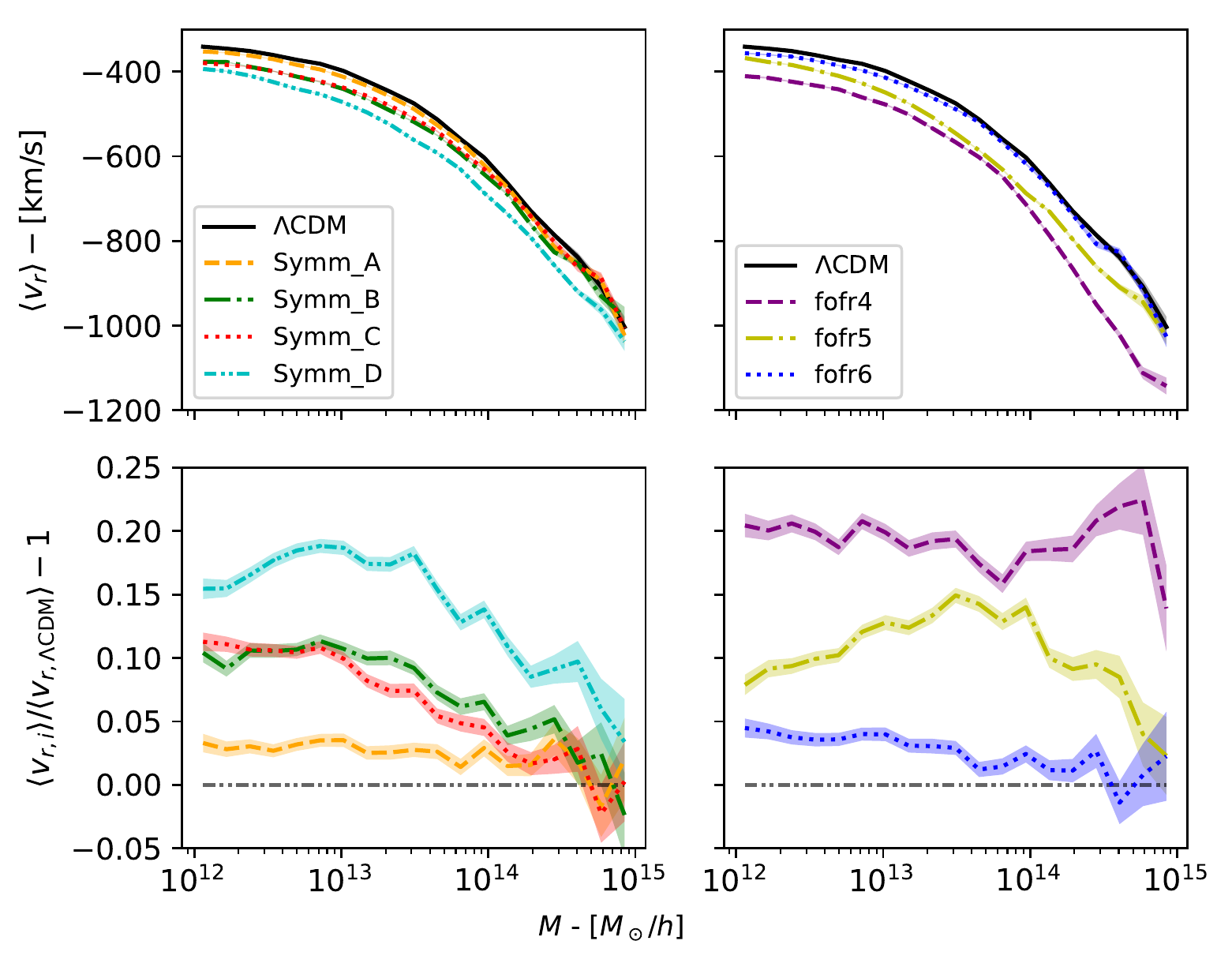}
\caption{The average radial speed of the filaments within a given mass bin. The negative value of the radial speed indicates that the particles are moving toward the filament. As expected, the particles moves faster inward the filament as the mass increases.}
\label{fig:Average_radial_speed_mass_bins}
\end{figure}

In this section, we show the average radial and tangential speeds of different mass ranges. The average radial speed within a mass bin is shown in the top row of Fig. \ref{fig:Average_radial_speed_mass_bins}, with the relative differences with respect to $\Lambda$CDM in the bottom row. The radial component is defined such that a positive value indicates that particles are moving outward the filaments. As expected, we see that the particles are, on average, moving toward the filament. The increase in radial speed due to an increased mass is caused by the fact that the gravitational potentials are deeper for the more massive filaments. The relative differences can also be compared to the ones in Fig. \ref{fig:Average_Speed_mass_bins}, and it is easy to see that the relative differences between the two figures are almost the same. This indicates that the radial speed will make up the majority of the total speed. 

\begin{figure}
\includegraphics[width=\linewidth]{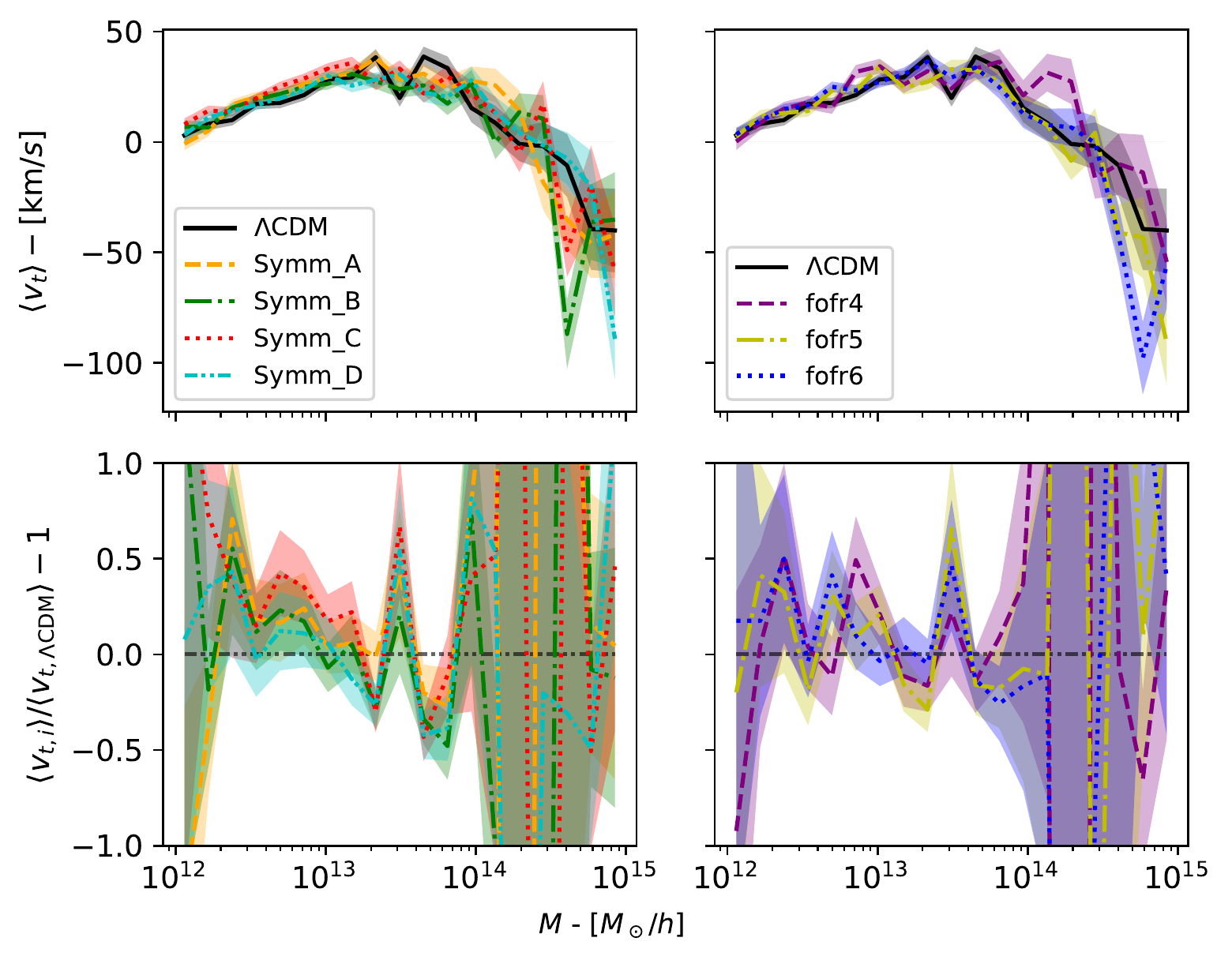}
\caption{The average tangential speed of the filaments within a given mass bin. }
\label{fig:Average_tangetial_speed_mass_bins}
\end{figure}

Figure \ref{fig:Average_tangetial_speed_mass_bins} shows the average tangential speed at different mass bins. For the least massive filaments, the particles generally have no preferential direction along the filament axis. For the intermediate massed filaments, the particles tend to move toward the endpoint which is considered to be the maximum point. Meanwhile, for the most massive filament, most particles appears to move toward the endpoint which defines the two-saddle. We found that most maximum points resided within a halo defined by the AHF halo catalog, while most of the two-saddles did not. The maximum points are therefore more dense. However, it is puzzling that the particles in the most massive filaments tend to move toward the less dense endpoint.

\end{appendix}

\end{document}